\documentstyle[12pt,leqno]{article}

\font\fifteen=cmbx10 at 15pt
\font\twelve=cmbx10 at 12pt
\font\bfit=cmbxti10 at 12pt

\font\twelvegb=cmssbx10 at 12pt
\font\sevengb=cmssbx10 at 9pt
\font\fivegb=cmssbx10 at 7pt

\newfam\gba
\textfont\gba=\twelvegb
\scriptfont\gba=\sevengb
\scriptscriptfont\gba=\fivegb
\def\gb{\fam\gba\twelvegb}

\newcommand{\bthe}{\mbox{\boldmath $\theta$}}
\newcommand{\bun}{\mbox{\boldmath $1$}}

\font\twelvegi=cmmib10 at 12pt
\font\sevengi=cmmib10 at 9pt
\font\fivegi=cmmib10 at 7pt

\newfam\git
\textfont\git=\twelvegi
\scriptfont\git=\sevengi
\scriptscriptfont\git=\fivegi
\def\gi{\fam\git\twelvegi}

\let\ul=\underline
\let\ol=\overline
\let\wt=\widetilde

\def\int{\mathop{\rm Int}\nolimits}
\def\Sup{\mathop{\rm Sup}\nolimits}
\def\tang{\mathop{\rm tang}\nolimits}

\newcommand{\be}[1]{\begin{equation}\label{#1}}
\newcommand{\ee}{\end{equation}}
\newcommand{\ba}{\begin{array}}
\newcommand{\ea}{\end{array}}
\newcommand{\bea}[1]{\begin{eqnarray}\label{#1}}
\newcommand{\eea}{\end{eqnarray}}

\newcommand{\rf}[1]{(\ref{#1})}
\newcommand{\rs}[1]{{\large\bf\ref{#1}}}

\newcommand{\dis}{\displaystyle}

\newcounter{saveeqn}
\newcommand{\alpheqn}{\setcounter{saveeqn}{\value{equation}}%
\stepcounter{saveeqn}\setcounter{equation}{0}%
\renewcommand{\theequation}{\mbox{\arabic{saveeqn}\alph{equation}}}}
\newcommand{\reseteqn}{\setcounter{equation}{\value{saveeqn}}%
\renewcommand{\theequation}{\arabic{equation}}}

\def\thebibliography#1{\section*{Bibliography}\list
 {[\arabic{enumi}]}{\settowidth\labelwidth{[#1]}\leftmargin\labelwidth
 \advance\leftmargin\labelsep
 \usecounter{enumi}}
 \def\newblock{\hskip .11em plus .33em minus .07em}
 \sloppy\clubpenalty4000\widowpenalty4000
 \sfcode`\.=1000\relax}

\begin{document}

\begin{titlepage}

\begin{center}

\renewcommand{\thefootnote}{\fnsymbol{footnote}}

{\twelve Centre de Physique Th\'eorique\footnote{
Unit\'e Propre de Recherche 7061
} - CNRS - Luminy, Case 907}

{\twelve F-13288 Marseille Cedex 9 - France }

\vspace{2 cm}

{\fifteen A DETAILED ACCOUNT OF ALAIN CONNES' VERSION \\
OF THE STANDARD MODEL IV}

\vspace{0.3 cm}

\setcounter{footnote}{0}
\renewcommand{\thefootnote}{\arabic{footnote}}

{\bf Daniel KASTLER
\footnote{ and Universit\'e d'Aix-Marseille II} \\
 and Thomas SCH\"UCKER
\footnote{ and Universit\'e de Provence} \\}

\vspace{4,5 cm}

{\bf Abstract}

\end{center}

We give a detailed account of the computation of the Yang-Mills action
for the Connes-Lott model with general coupling constant in the
commutant of the $K$-cycle. This leads to tree-approximation results
amazingly compatible with experiment, yielding a first indication on
the Higgs mass.

\vspace{4,5 cm}

\noindent PACS-92: 11.15 Gauge field theories

\noindent MSC-91: 81E13 Yang-Mills and other gauge theories

\bigskip

\noindent January 1995

\noindent CPT-94/P.3092

\noindent hep-th/9501077

\bigskip

\noindent anonymous ftp or gopher: cpt.univ-mrs.fr

\end{titlepage}

The preceding papers I, II and III of this series presented
computations based on the following Ansatz for the non-commutative
Yang-Mills action:
\be{1}
Y M = \alpha_l \left( \bthe_l, \bthe_l \right)_l
+ \alpha_q \left( \bthe_q, \bthe_q \right)_q
\kern 1,5cm  \left( \alpha_l + \alpha_q = 1 \right),
\ee
with the scalar products
$(\cdot \, , \cdot)_l$ and
$(\cdot \, , \cdot)_q$
stemming as follows from the traces
$\tau_{{\gi D}_l}$ and
$\tau_{{\gi D}_q}$:
\be{2}
\left\{
\ba{ll}
(\omega, \omega')_l & = Re \tau_{{\gi D}_l}
\left( \omega^{\star} \omega' \right) = Re Tr_{\omega}
\left\{
{\gi D}_l^{- 4} \pi_l \left( \omega^{\star} \omega' \right)
\right\} \\[2mm]
(\omega, \omega')_q & = Re \tau_{{\gi D}_q}
\left( \omega^{\star} \omega' \right) = Re Tr_{\omega}
\left\{
{\gi D}_q^{- 4} \pi_q \left( \omega^{\star} \omega' \right)
\right\}
\ea
\right. ,
\kern 4mm
\omega , \omega' \in \Omega {\gb A}^n (\in \Omega {\gb B}^n).
\ee

These scalar products are however not the most general
and natural ones~\cite{3b}: with
$\ul \Gamma_l$, resp.
$\ul \Gamma_q$ ($\ul \Gamma'_l$, resp.
$\ul \Gamma'_q$), positive
elements of the respective commutants
$\{ \pi_l ({\gb A}), {\gi D}_l \}'$,
$\{ \pi_q ({\gb A}), {\gi D}_q \}'$
$\left( \{ \pi_l ({\gb B}), {\gi D}_l \}' \right.$,
$\left. \{ \pi_q ({\gb B}), {\gi D}_q \}' \right)$, the alternative
Ansatz:
  \alpheqn
\be{3a}
\left\{
\ba{ll}
(\omega, \omega')_l & = Re Tr_{\omega}
\left\{
{\gi D}_l^{- 4} \pi_l \left( \omega^{\star} \omega' \right)
\ul \Gamma_l
\right\} \\[2mm]
(\omega, \omega')_q & = Re Tr_{\omega}
\left\{
{\gi D}_q^{- 4} \pi_q \left( \omega^{\star} \omega' \right)
\ul \Gamma_q
\right\}
\ea
\right. ,
\kern 4mm
\omega , \omega' \in \Omega {\gb A}^n,
\ee
\be{3b}
\left\{
\ba{ll}
(\omega, \omega')_l & = Re Tr_{\omega}
\left\{
{\gi D}_l^{- 4} \pi_l \left( \omega^{\star} \omega' \right)
\ul \Gamma'_l
\right\} \\[2mm]
(\omega, \omega')_q & = Re Tr_{\omega}
\left\{
{\gi D}_q^{- 4} \pi_q \left( \omega^{\star} \omega' \right)
\ul \Gamma'_q
\right\}
\ea
\right. ,
\kern 4mm
\omega , \omega' \in \Omega {\gb B}^n,
\ee
  \reseteqn
yield indeed, in contrast to the previous  Ansatz~\rf{2}, a
non-committed choice of (as required) gauge-invariant scalar products.
According to the Poincar\'e-duality philosophy the new Yang-Mills
action will be the sum of its electroweak and chromodynamics parts
respectively stemming from the scalar products~\rf{3a} and
\rf{3b}.\footnote{
The scalar products~\rf{3a}, \rf{3b} incorporate a
generalized coupling constant ``in the commutant of the
$K$-cycle''. The leptonic and quark sectors appear independently since
their intertwiners are trivial cf.~\rf{46}, \rf{47} below.
} Due to the product structure of the traces
$\tau_{{\gi D}_l}$ and
$\tau_{{\gi D}_q}$:\footnote{
$Tr_2$ denotes the
$R - L\ 2 \times 2$ matrix trace
with entries matrices given by the weak isotopic spin (overall
$4 \times 4$ matrices for quarks, cf.~\rf{16}, \rf{17}, \rf{18}
below; and
$3 \times 3$ matrices for leptons, cf.~\rf{19}, \rf{20}, \rf{21}
below; whilst
$Tr_N$,
$Tr_3$, and
$\tau_D$ respectively stand for
the fermion family
$N \times N$ matrix trace, the colour
$3 \times 3$ matrix trace, and the Dirac-operator trace
$\tau_D$.
}
\be{4}
\left\{
\ba{ll}
\tau_{{\gi D}_l} & = \tau_D \otimes Tr_2 \otimes Tr_N \\[2mm]
\tau_{{\gi D}_q} & = \tau_D \otimes Tr_2 \otimes Tr_N \otimes Tr_3
\ea
\right.
,
\ee
together with the ``fiberwise" nature of the combined
space-time-inner space theory, the problem of adapting our previous
computations to the generalized Ansatz~\rf{3a}, \rf{3b} will de facto
reduce to computations within the representations
$\pi_l$ and
$\pi_q$
pertaining to the inner space. The present treatment differs also from
that of III in that the modular coalescence of the three
$U (1)$ groups -- and their Lie algebras is performed after squaring
the curvature and not before.

The results reached through the new computations constitute a
relatively mild modification of our previous results based on the
former Ansatz~\rf{2}, however leading to much more satisfying
tree-approximation results, suppressing the former
inconsistencies\footnote{
e.g. the fact that, for
$x = 0$, the values of
$g_3 / g_2 = 1$ and
$\sin^2 \theta = 3 / 8$ were of the ``grand unification type"
whilst the mass-ratio
$m_t / m_W = 2$ was near the experimental value~\cite{5b}.
} and amazingly compatible with the experimental evidence. For a
specific choice of the ``coupling constant'' within the commutant,
one computes the tree-approximation values of the ratio between
strong and electroweak coupling constant, the weak angle, and the
ratios between the top and the
$W$, and the Higgs and the top masses.
One can fit the three first items with the known experimental values:
in fact
$g_3 / g_2$ and
$\sin^2 \theta_W$ turn out to be mutually
uncorrelated, and uncorrelated with the ratios
$m_t / m_W$ and $m_H / m_W$ which determine each other, thus yielding,
since the top has been found, a ``prediction'' of the Higgs mass: the
latter equals
$1.5698 \ m_t$ for $m_t / m_W = 2$, the value (near experiment~!)
fixed by the (canonical~?) choice of the ``coupling constant'' in the
center of the
$K$-cycle, a choice without incidence on
$g_3 / g_2$ and
$\sin^2 \theta_W$.

Of course these results are ``classical''. Reliable results await a
renormalized field quantization which should be effected with due
consideration of the (still to be found) esoteric symmetry brought
about by the Higgs boson as a fifth gauge boson.

\vspace{6mm}

This paper is the companion paper to~\cite{5a} with whom it shares
its subject matter with a different style and emphasis.
Whilst~\cite{5a}, destined to a physical audience, adopts a
notational setting congenial to the habits of elementary particle
physicists, and insists on the global strategy~\cite{6} without
giving all computational details, we here address mathematical
physicists in the notation of our former reports~\cite{4b}, giving a
line-by-line account of computations. These computations have been
performed independently in the two papers in different notation, thus
affording a mutual check.

\vspace{6mm}

For the convenience of our reader, we begin by recalling the
definitions of the inner space structure with its Poincar\'e-dual
${\bf A}_{\rm ew} \otimes {\bf B}_{\rm chrom}$-$K$-cycles
$(\ul H_l, \ul D_l, \ul \chi_l)$ and
$(\ul H_q, \ul D_q, \ul \chi_q)$. We begin with the
uncolored leptonic and quarkonic
${\bf A}_{\rm ew}$-$K$-cycles.

\vspace{4mm}


\setcounter{section}{-1}

\section{Reminder (the inner space).}\label{s0}

{\bfit The algebra}
${\bf A}_{\rm ew}$. With
${\gi H} = \left\{
\normalbaselineskip=18pt
\pmatrix{
a       & b     \cr
- \ol b & \ol a \cr
}
; a, b \in {\gi C}
\right\}$, we have:
\be{5}
{\bf A}_{\rm ew} = {\gi C} \oplus {\gi H} =
\{ (p,q) ; p \in {\gi C}, q \in {\gi H} \} =
\left\{
\left(
\normalbaselineskip=18pt
\pmatrix{
\ol p & 0 \cr
0     & p \cr
}
, q
\right) ; p \in {\gi C}, q \in {\gi H}
\right\}
\ee
\be{6}
(p, q)^{\star} = \left( \ol p, q^{\star} \right),
\kern 2cm  p \in {\gi C},\ q \in {\gi H},
\ee
\be{7}
{\bf G}_{\rm ew} =
\left\{
{\gi u} = (u, v) \in {\bf A}_{\rm ew} ;
u^{\star} u = u u^{\star} = \bun, v^{\star} v = v v^{\star} = \bun
\right\}
= U (1) \times SU (2)
\ee

\noindent {\bfit The leptonic and quarkonic
${\bf A}_{\rm ew}$-K-cycles
$(H_l, D_l, \chi_l)$ and
$(H_q, D_q, \chi_q)$.
}

\noindent {\bfit Leptonic K-cycle:} Hilbert space:
\be{8}
\ba{ll}
H_l = & \left( {\gi C}_R^1 \oplus {\gi C}_L^2 \right)
\otimes {\gi C}^N ,                              \\[2mm]
      & \kern 3mm  e_R  \kern 5mm
\nu_L e_L
\ea
\ee
Operators (endomorphisms of
$H_l$ as
$3 \times 3$ matrices with entries in
$M_N ({\gi C})$):
\be{9}
\chi_l = \kern -2mm
\bordermatrix{
& e_R    & \nu_L    & e_L      \cr
& \bun_N & 0        & 0        \cr
& 0      & - \bun_N & 0        \cr
& 0      & 0        & - \bun_N \cr
}
\ba{l}
e_R   \\
\nu_L \\
e_L
\ea
{}.
\ee
\be{10}
\pi_l ((p, q)) = \kern -2mm
\bordermatrix{
& e_R      & \nu_L          & e_L          \cr
& p \bun_N & 0              & 0            \cr
& 0        & a \bun_N       & b \bun_N     \cr
& 0        & - \ol b \bun_N & \ol a \bun_N \cr
}
\ba{l}
e_R   \\
\nu_L \\
e_L
\ea
, \kern 2mm
\left(
p =
\normalbaselineskip=18pt
\pmatrix{
\ol p & 0 \cr
0     & p \cr
},
q =
\normalbaselineskip=18pt
\pmatrix{
a       & b     \cr
- \ol b & \ol a \cr
}
\right)
\in {\bf A}_{\rm ew},
\ee
\be{11}
D_l = \kern -2mm
\bordermatrix{
& e_R & \nu_L & e_L         \cr
& 0   & 0     & M_e^{\star} \cr
& 0   & 0     & 0           \cr
& M_e & 0     & 0           \cr
}
\ba{l}
e_R   \\
\nu_L \\
e_L
\ea
{}.
\ee

\noindent {\bfit Quarkonic K-cycle:} Hilbert space:
\be{12}
\ba{ll}
H_q = & \left( {\gi C}_R^2 \oplus {\gi C}_L^2 \right)
\otimes {\gi C}^N ,                             \\[2mm]
      & \mkern 1mu  u_R d_R  \kern 3mm
u_L d_L  \kern 3mm
\ea
\ee
Operators (endomorphisms of
$H_q$ as
$4 \times 4$ matrices with entries in
$M_N ({\gi C})$):
\be{13}
\chi_q = \kern -2mm
\bordermatrix{
& u_R    & d_R      & u_L      & d_L      \cr
& \bun_N & 0        & 0        & 0        \cr
& 0      & \bun_N   & 0        & 0        \cr
& 0      & 0        & - \bun_N & 0        \cr
& 0      & 0        & 0        & - \bun_N \cr
}
\ba{l}
u_R \\
d_R \\
u_L \\
d_L
\ea
{}.
\ee
\be{14}
\kern 6mm  \pi_q ((p, q)) =  \kern -2mm
\bordermatrix{
& u_R          & d_R      & u_L            & d_L          \cr
& \ol p \bun_N & 0        & 0              & 0            \cr
& 0            & p \bun_N & 0              & 0            \cr
& 0            & 0        & a \bun_N       & b \bun_N     \cr
& 0            & 0        & - \ol b \bun_N & \ol a \bun_N \cr
}
\ba{l}
u_R \\
d_R \\
u_L \\
d_L
\ea
, \kern 2mm
\left(
p =
\normalbaselineskip=18pt
\pmatrix{
\ol p & 0 \cr
0     & p \cr
},
q =
\normalbaselineskip=18pt
\pmatrix{
a       & b     \cr
- \ol b & \ol a \cr
}
\right)
\in {\bf A}_{\rm ew},
\ee
\be{15}
D_q = \kern -2mm
\bordermatrix{
& u_R & d_R & u_L         & d_L         \cr
& 0   & 0   & M_u^{\star} & 0           \cr
& 0   & 0   & 0           & M_d^{\star} \cr
& M_u & 0   & 0           & 0           \cr
& 0   & M_d & 0           & 0           \cr
}
\ba{l}
u_R \\
d_R \\
u_L \\
d_L
\ea
,
\ee

\noindent {\it Remark:} One passes from the
matrices~\rf{5}, \rf{6}, \rf{7} to the
matrices~\rf{8}, \rf{9}, \rf{10} through the changes
$M_u \to 0$, $M_d \to M_e$ followed by restriction to the right-lower
corner
$3 \times 3$ matrix. This procedure applied to a
$3 \times 3$
depending upon
$M_u$ and
$M_d$ is called {\bf leptonic reduction}.

We will in fact also use the following

\noindent {\bfit Two-by-two matrix versions. Quark sector:} version
with
$2 \times 2$ matrices with entries in
$M_2 ({\gi C}) \otimes M_N ({\gi C})$, corresponding to the
decomposition:
\be{16}
\pi_q ((p, q)) = \kern -2mm
\bordermatrix{
& R                & L                \cr
& p \otimes \bun_N & 0                \cr
& 0                & q \otimes \bun_N \cr
}
\ba{l}
R \\
L
\ea
,
\ee
\be{17}
D_q = \kern -2mm
\bordermatrix{
& R       & L               \cr
& 0       & {\gi M}^{\star} \cr
& {\gi M} & 0               \cr
}
\ba{l}
R \\
L
\ea
,
\ee
\be{18}
\chi_q = \kern -2mm
\bordermatrix{
& R                   & L                     \cr
& \bun \otimes \bun_N & 0                     \cr
& 0                   & - \bun \otimes \bun_N \cr
}
\ba{l}
R \\
L
\ea
{}.
\ee

\noindent {\bfit Lepton sector:} version with
$2 \times 2$ matrices with entries
$$
\normalbaselineskip=18pt
\pmatrix{
 M_1 \otimes M_N ({\gi C})
& M ({\gi C}^2,{\gi C}) \otimes M_N ({\gi C}) \cr
 M ({\gi C},{\gi C}^2) \otimes M_N ({\gi C})
& M_2 ({\gi C}) \otimes M_N ({\gi C})         \cr
}:
$$
\be{19}
\kern 6mm  p_l ((p, q)) =  \kern -2mm
\bordermatrix{
& R                & L                \cr
& p \otimes \bun_N & 0                \cr
& 0                & q \otimes \bun_N \cr
}
\ba{l}
R       \\
L
\ea
, \kern 3mm
\left(
p \in {\gi C},\ q =
\normalbaselineskip=18pt
\pmatrix{
a       & b     \cr
- \ol b & \ol a \cr
}
\in {\bf A}_{\rm ew}
\right)
,
\ee
\be{20}
D_l = \kern -2mm
\bordermatrix{
& R          & L        \cr
& 0          & (0\ M_e^{\star}) \cr
& \normalbaselineskip=12pt
\pmatrix{
0   \cr
M_e \cr
}            & 0                \cr
}
\ba{l}
R \\
L
\ea
{}.
\ee
\be{21}
\chi_l = \kern -2mm
\bordermatrix{
& R              & L                     \cr
& \bun \otimes 1 & 0                     \cr
& 0              & - \bun \otimes \bun_N \cr
}
\ba{l}
R \\
L
\ea
{}.
\ee

\noindent {\bfit Coloured Poincar\'e-dual
${\bf A}_{\rm ew} \otimes {\bf B}_{\rm chrom}$-K-cycles
$\left( \ul H_l, \ul D_l, \ul \chi_l \right)$ and
$\left( \ul H_q, \ul D_q, \ul \chi_q \right)$.

\noindent The algebra
${\bf B}_{\rm chrom}$.
}
\be{22}
{\bf B}_{\rm chrom} = {\gi C} \oplus M_3 ({\gi C}) =
\left\{
(p', M) ; p \in {\gi C},\ m \in M_3 ({\gi C})
\right\}
\ee
\be{23}
(p', m)^{\star} = \left( \ol p', m^{\star} \right),
\kern 2cm  p' \in {\gi C},\ m \in M_3 ({\gi C}),
\ee
\be{24}
{\bf G}_{\rm chrom} =
\left\{
{\gi u}' = (u', v) \in {\bf A}_{\rm ew} ;
u^{\star} u = u u^{\star} = \bun, v^{\star} v = v v^{\star} = \bun
\right\}
= U (1) \times U (3).
\ee

\noindent {\bfit The leptonic and quarkonic
${\bf A}_{\rm ew} \otimes {\bf B}_{\rm chrom}$-K-cycles
$(\ul H_l, \ul D_l, \ul \chi_l)$ and
$(\ul H_q, \ul D_q, \ul \chi_q)$.

\noindent Leptonic K-cycle
$(\ul H_l, \ul D_l, \ul \chi_l)$:
}
\be{25}
\left\{
\ba{l}
  \ul H_l = H_l \otimes {\gi C}_{\rm chrom},  \kern 1cm
\ul \chi_l = \chi_l \otimes \bun_{\rm chrom}               \\[2mm]
  \ul D_l = D_l \otimes \bun_{\rm chrom}                   \\[2mm]
  \ul \pi_l (p, q) = \pi_l (p, q) \otimes \bun_{\rm chrom} \\[2mm]
\ul \pi_l (p', m) = \bun_l \otimes p' = p'
\ea
\right. .
\ee

\noindent {\bfit Quarkonic K-cycle
$(\ul H_q, \ul D_q, \ul \chi_q)$:
}
\be{26}
\left\{
\ba{l}
  \ul H_q = H_q \otimes {\gi C}_{\rm chrom}^3 \, , \kern 1cm
\ul \chi_q = \chi_q \otimes \bun_{\rm chrom}               \\[2mm]
  \ul D_q = D_q \otimes \bun_{\rm chrom}                   \\[2mm]
  \ul \pi_q (p, q) = \pi_q (p, q) \otimes \bun_{\rm chrom} \\[2mm]
\ul \pi_q (p', m) = \bun_q \otimes m
\ea
\right.
\ee
(note the relation
$[\ul D_l, \ul \pi_l (p', m)]
= [\ul D_q, \ul \pi_q (p', m)] = 0$ implying the
algebraic Poincar\'e duality condition).

We recall the formulae (concerning the quark sector):
\be{27}
\left\{
\ba{l}
{\gi M} =
\normalbaselineskip=18pt
\pmatrix{
M_u & 0   \cr
0   & M_d \cr
}
= E \otimes M_u + F \otimes M_d \\[4mm]
\mbox{with}\ E =
\normalbaselineskip=18pt
\pmatrix{
1 & 0 \cr
0 & 0 \cr
},
\ \mbox{and}\ F =
\normalbaselineskip=18pt
\pmatrix{
0 & 0 \cr
0 & 1 \cr
}.
\ea
\right.
\ee

\bigskip

\noindent {\bf COMMUTANTS OF THE INNER SPACE ELECTROWEAK, RESP.
CHROMODYNAMICS ${\gi K}$-CYCLE.}

\medskip

We now analyze the {\it commutants of the
inner space electroweak, resp.chromodynamics K-cycle}, calling so the
respective subalgebras of End
$(\ul H_l \oplus \ul H_q)$ consisting of the elements
commuting with
$\ul D_l \oplus \ul D_q$ and with all
$\ul \pi_l (a) \oplus \pi_q (a),\ a \in {\bf A}_{\rm ew}$, resp. all
$\ul \pi_l (b) \oplus \pi_q (b),\ b \in {\bf B}_{\rm chrom}$. We begin
with a remark relative to a notation which we shall use in order to
spare writing:


\section{Remark.}\label{s1}

With
$a$ and
$b$ linear operators of the respective complex vector spaces
$H$ and
$K$, we write
$\int \, (a, b)$ for the set of linear maps:
$H \to K$
intertwining
$a$ and
$b$:
\be{28}
\int \, (a, b) =
\left\{
S \in \mbox{End}\ (H, K) ; S a = b S)
\right\}.
\ee
We then have that:

\begin{enumerate}

\item With
$a, H$ and
$b, K$ as above, and using a
$2 \times 2$ matrix notation for the endomorphisms of
$H \oplus K$, we have that:
\be{29}
\int \, (a \oplus b, a \oplus b) =
\normalbaselineskip=18pt
\pmatrix{
\int \, (a, a) & \int \, (a, b) \cr
\int \, (b, a) & \int \, (b, a) \cr
}.
\ee

\item With
$a, H$ and
$b, K$ as above,
$H$ and
$K$ finite-dimensional, and
$a$ and
$b$ self-adjoint with non-intersecting respective sets of eigenvalues,
we have
$\int \, (a, b) = \{ 0 \}$.

\end{enumerate}

\pagebreak

\noindent \ul{Proof:}

\begin{enumerate}

\item follows from:
\be{30}
\left[
\normalbaselineskip=18pt
\pmatrix{
a & b \cr
c & d \cr
},
\normalbaselineskip=18pt
\pmatrix{
S & 0 \cr
0 & T \cr
}
\right]
=
\normalbaselineskip=18pt
\pmatrix{
a S - S a & b T - S b \cr
c S - T c & d T - T d \cr
}.
\ee

\item With
$S = (S_k^i),\ a = (\lambda_i \delta_k^i),\ b = (\mu_i \delta_k^i)$,
we have
\be{31}
(S a - b S)_k^i = \Sigma_h
\left(
S_h^i \lambda_h \delta_k^h - \mu_i \delta_h^i S_k^h
\right)
= \left( \lambda_k - \mu_i \right) S_k^i = 0.
\ee

In what follows we shall comply to the common usage of choosing
our fermion mass-matrices such that
$M_e$ {\it and
$M_u$ are diagonal, positive matrices, whilst
$M_d = C | M_d |$, with}
$C$ (the {\bf Kobayashi-Maskawa matrix}) {\it unitary and
$| M_d |$ strictly positive}. Furthermore we assume that {\it all
fermion masses are different} (the eigenvalues of
$M_e$,
$M_u$ and
$| M_d |$ consists of positive numbers (the masses of leptons and
quarks) all different from one another -- experiment!). We further
assume that {\it no eigenstate of
$| M_d |$ is an  eigenstate of}
$C$ (experiment!). We use the shorthands:
\be{32}
\left\{
\ba{cll}
\mu   & = {\gi M} {\gi M}^{\star}, \kern 1cm \wt \mu
      & = {\gi M}^{\star} {\gi M},                     \\[2mm]
\mu_e & = M_e^2                                        \\[2mm]
\mu_u & = M_u^2                                        \\[2mm]
\mu_d & = M_d M_d^{\star}, \kern 1cm \wt \mu_d
          & = M_d^{\star} M_d,
\ea
\right.
\ee
we then have:
\be{33}
\wt \mu_d = | M_d |^2 = C \mu_d C^{\star},
\ee
 \setcounter{equation}{32}
  \alpheqn
\be{33a}
\wt \mu = | {\gi M} |^2 = C \mu C^{\star},
\ee
  \reseteqn
where
$C = i d \oplus C$ in the second line.

\end{enumerate}


\section{Lemma.}\label{s2}

{\it We have that:

\begin{enumerate}

\item The most general self-adjoint element
$\Gamma'_l$ of
$\int \, (D_l, D_l)$ is as follows: one has in
$3 \times 3$ matrix notation:
\be{34}
\Gamma'_l =
\normalbaselineskip=18pt
\pmatrix{
h (\mu_e) & 0      & k (\mu_e) \cr
0         & \delta & 0         \cr
k (\mu_e) & 0      & h (\mu_e) \cr
},
\ee
where
$h$ and
$k$ are arbitrary real functions, and
$\delta$ is any self-adjoint element of
$M_N ({\gi C})$.

\item The most general self-adjoint element
$\Gamma'_q$ of
$\int \, (D_q, D_q)$ is as follows: one has in
$4 \times 4$ matrix notation:
\be{35}
\Gamma'_q =
\normalbaselineskip=18pt
\pmatrix{
f (\mu_u) & 0 & l (\mu_u) & 0 \cr
0 & g \left( \wt \mu_d \right) &
0 & m \left( \wt \mu_d \right) C^{\star} \cr
l (\mu_u) & 0 & f (\mu_u) & 0 \cr
0 & C m \left( \wt \mu_d \right) &
0 & C g \left( \wt \mu_d \right) C^{\star} \cr
},
\ee
where
$f, g, l$ and
$m$ are arbitrary real functions. Thus the most general self-adjoint
element
$\ul \Gamma'_q$ of
$\int \, (\ul D_q, \ul D_q)$ is of the form
$\Gamma'_q \otimes S$ with
$\Gamma'_q$ as in~\rf{35} and
$S \in M_N ({\gi C})$ self-adjoint.

\item The self-adjoint elements of
$\int \, (D_l, D_q)$ or of
$\int \, (D_q, D_l)$ vanish. The same holds for self-adjoint elements
of
$\int \, (\ul D_l, \ul D_q)$ or of
$\int \, (\ul D_q, \ul D_l)$.

\end{enumerate}
}

\noindent \ul{Proof:} \begin{enumerate}

\item With
$\Gamma'_l =
\normalbaselineskip=18pt
\pmatrix{
\alpha    & \beta   & \mu    \cr
\ol \beta & \delta  & \nu    \cr
\ol \mu   & \ol \nu & \sigma \cr
}$, in
$3 \times 3$ matrix notation, equating:
\be{36}
D_l \Gamma'_l =
\normalbaselineskip=18pt
\pmatrix{
0   & 0 & M_e \cr
0   & 0 & 0   \cr
M_e & 0 & 0   \cr
}
\normalbaselineskip=18pt
\pmatrix{
\alpha    & \beta   & \mu    \cr
\ol \beta & \delta  & \nu    \cr
\ol \mu   & \ol \nu & \sigma \cr
} =
\normalbaselineskip=18pt
\pmatrix{
M_e \ol \mu & M_e \ol \nu & M_e \sigma \cr
0           & 0           & 0          \cr
M_e \alpha  & M_e \beta   & M_e \mu    \cr
}
\ee
and
\be{37}
\Gamma'_l D_l =
\normalbaselineskip=18pt
\pmatrix{
\alpha    & \beta   & \mu    \cr
\ol \beta & \delta  & \nu    \cr
\ol \mu   & \ol \nu & \sigma \cr
}
\normalbaselineskip=18pt
\pmatrix{
0   & 0 & M_e \cr
0   & 0 & 0   \cr
M_e & 0 & 0   \cr
} =
\normalbaselineskip=18pt
\pmatrix{
\mu M_e    & 0 & \alpha M_e    \cr
\nu M_e    & 0 & \ol \beta M_e \cr
\sigma M_e & 0 & \ol \mu M_e   \cr
},
\ee
yields
$\beta = \nu = 0$; further
$M_e \ol \mu = \mu M_e$ and
$M_e \mu = \ol \mu M_e$, whence
$M_e^2 \ol \mu = M_e \mu M_e = \ol \mu M_e^2$, whence
$\ol \mu = k (M_e)$,
$\mu = k' (M_e)$ for some functions
$k, k'$, with in addition
$k = k'$ since the relation
$M_e \ol \mu = \mu M_e$ now reads
$M_e \Bigl( k (M_e) - k' (M_e) \Bigr) = 0$; finally
$M_e \alpha = \sigma M_e$ and
$M_e \sigma = \alpha M_e$, whence
$M_e^2 \alpha = M_e \sigma M_e = \alpha M_e^2$, whence
$\alpha = h \left( M_e^2 \right)$ which also equals
$\sigma$ because the relation
$M_e \alpha = \sigma M_e$ now reads
$\alpha M_e = \sigma M_e$.

\item We have, using the
$2 \times 2$ matrix notation, with
$\Gamma'_q =
\normalbaselineskip=18pt
\pmatrix{
a & b \cr
c & d \cr
}$,
$a, b, c, d \in M_2 ({\gi C}) \otimes M_N ({\gi C})$:
\be{38}
\left[ \Gamma'_q, D_q \right] =
\left[
\normalbaselineskip=18pt
\pmatrix{
a & b \cr
c & d \cr
},
\normalbaselineskip=18pt
\pmatrix{
0       & {\gi M}^{\star} \cr
{\gi M} & 0               \cr
}
\right]
=
\normalbaselineskip=18pt
\pmatrix{
 b {\gi M} - {\gi M}^{\star} c
& a {\gi M}^{\star} - {\gi M}^{\star} d \cr
 d {\gi M} - {\gi M} a
& c {\gi M}^{\star} - {\gi M} b         \cr
},
\ee
we must thus have:
  \alpheqn
\bea{39a}
& & d {\gi M} - {\gi M} a = 0,                             \\[2mm]
& & a {\gi M}^{\star} - {\gi M}^{\star} d = 0, \label{39b} \\[2mm]
& & b {\gi M} - {\gi M}^{\star} c = 0,         \label{39c} \\[2mm]
& & c {\gi M}^{\star} - {\gi M} b = 0,         \label{39d}
\eea
  \reseteqn
We compute
$a$ and
$d$:~\rf{39a} and \rf{39b} entail:
\be{40}
a {\gi M}^{\star} {\gi M} = {\gi M}^{\star} {\gi M} a, \qquad
d {\gi M} {\gi M}^{\star} = {\gi M} {\gi M}^{\star} d.
\ee
whence the existence of a real function
$F$ with
$a = F \left( \wt \mu \right)$.\rf{39a} then reads
$d C | {\gi M} | = C | {\gi M} | F \left( \wt \mu \right)
= C F \left( \wt \mu \right) | {\gi M} |$ whence
$d C = C F \left( \wt \mu \right)$,
$d = C F \left( \wt \mu \right) C^{\star}$.

We now compute
$b$ and
$c$:~\rf{39c} and \rf{39d} entail
${\gi M}^{\star} {\gi M} b = {\gi M}^{\star} c {\gi M}^{\star}
= b {\gi M} {\gi M}^{\star}$,
$\wt \mu b = b \mu = b C \wt \mu C^{\star}$,
$\wt \mu b C = b C \wt \mu$, whence the existence of a real function
$G$ with
$b = G \left( \wt \mu \right) C^{\star}$. For
$c = b^{\star}$,
$G$ is  real: indeed~\rf{39d} then reads
${\gi M} b = C | {\gi M} | b = b^{\star} {\gi M}^{\star} =
b^{\star} | {\gi M} | C^{\star}$, i.e.
$| {\gi M} | b C = C^{\star} b^{\star} | {\gi M} |$, whence, since
$| {\gi M} | b C = b C | {\gi M} |$,
$b C = C^{\star} b^{\star}$.

We showed the existence of real functions
$F$ and
$G$ for which:
\be{41}
\Gamma'_q =
\normalbaselineskip=18pt
\pmatrix{
F \left( \wt \mu \right)   & G \left( \wt \mu \right) C^{\star}   \cr
C G \left( \wt \mu \right) & C F \left( \wt \mu \right) C^{\star} \cr
},
\ee
which comes to~\rf{35} in $4 \times 4$ matrix notation.

\item Let $S =
\normalbaselineskip=18pt
\pmatrix{
\alpha   & \beta   & \gamma   & \delta   \cr
\alpha'  & \beta'  & \gamma'  & \delta'  \cr
\alpha'' & \beta'' & \gamma'' & \delta'' \cr
} \in \int (D_q, D_l)$: equating

\be{42}
\normalbaselineskip=18pt
\pmatrix{
0   & 0 & M_e \cr
0   & 0 & 0   \cr
M_e & 0 & 0   \cr
}
\normalbaselineskip=18pt
\pmatrix{
\alpha   & \beta   & \gamma   & \delta   \cr
\alpha'  & \beta'  & \gamma'  & \delta'  \cr
\alpha'' & \beta'' & \gamma'' & \delta'' \cr
} =
\normalbaselineskip=18pt
\pmatrix{
M_e \alpha'' & M_e \beta'' & M_e \gamma'' & M_e \delta'' \cr
0            & 0           & 0            & 0            \cr
M_e \alpha   & M_e \beta   & M_e \gamma   & M_e \delta   \cr
}
\ee
and
\be{43}
\normalbaselineskip=18pt
\pmatrix{
\alpha   & \beta   & \gamma   & \delta   \cr
\alpha'  & \beta'  & \gamma'  & \delta'  \cr
\alpha'' & \beta'' & \gamma'' & \delta'' \cr
}
\normalbaselineskip=18pt
\pmatrix{
0   & 0   & M_u & 0           \cr
0   & 0   & 0   & M_d^{\star} \cr
M_u & 0   & 0   & 0           \cr
0   & M_d & 0   & 0           \cr
} =
\normalbaselineskip=18pt
\pmatrix{
\gamma M_u   & \delta M_d   & \alpha M_u   & \beta M_d^{\star}   \cr
\gamma' M_u  & \delta' M_d  & \alpha' M_u  & \beta' M_d^{\star}  \cr
\gamma'' M_u & \delta'' M_d & \alpha'' M_u & \beta'' M_d^{\star} \cr
}
\ee
yields
$\alpha' = \beta' = \gamma' = \delta' = 0$; further
$M_e \alpha'' = \gamma M_u$ and
$M_e \gamma = \alpha'' M_u$ whence
$M_e^2 \gamma = M_e \alpha'' M_u = \gamma M_u^2$ implying
$\gamma = 0$ by~\rs{s1}(ii), and thus
$\alpha'' = 0$ -- the changes
$\alpha'' \to \gamma''$ and
$\gamma \to \alpha$ then  yielding
$\alpha = \gamma'' = 0$; analogously
$M_e \beta'' = \delta M_d$ and
$M_e \delta = \beta'' M_d^{\star}$ whence
$M_e^2 \beta'' = M_e \delta M_d = \beta'' M_d^{\star} M_d$ implying
$\beta'' = 0 = \delta$; the changes
$\beta'' \to \delta''$ and
$\delta \to \beta$ then yielding
$\beta = \delta'' = 0$. We proved that
$S = 0$. For
$S \in \int (\ul D_q, \ul D_l)$, writing
$\ul H_q = H_q \otimes {\gi C}^3$ as a direct sum, the restrictions
of
$S$ to all summands vanish.

\end{enumerate}


\section{Lemma.}\label{s3}

{\it We have that:

\begin{enumerate}

\item The commutant of the $K$-cycle
$(H_q, D_q, \chi_q)$ of
${\bf A}_{\rm ew}$\footnote{
We recall that we call commutant of the
$K$-cycle
$(H, D, \chi)$ of an algebra
${\bf A}$ the set of operators of
$H$ commuting with
$\pi ({\bf A})$ and
$D$.
} coincides with
$\bun_2 \otimes \bun_N \otimes M_3 ({\gi C})$.\footnote{
$\bun_2$ denotes here the unit of the
$2 \times 2$ matrix algebra with entries in
$M_2 ({\gi C})$. Note that in fact the colorless
$K$-cycle
$(H_q, D_q, \chi_q)$ is irreducible in the sense that the only
operators of
$H_q$ commuting with
$\pi_q ({\bf A}_{\rm ew})$ and
$D_q$ are the scalars. Equivalently
$\pi_q ({\bf A}_{\rm ew})$ and
$D_q$ generate
$B (H_q)$.
}

\item A self-adjoint
$\ul \Gamma_l$ acting on
$\ul H_l$ belongs to the commutant of the
$K$-cycle
$(\ul H_l, \ul D_l, \ul \chi_l)$
of
${\bf A}_{\rm ew}$ iff it is of the type
$\bun_2 \otimes \Gamma_N$ , with
$\Gamma_N$ a real function of
$M_e$.\footnote{
$\bun_2$ denotes here the unit
$2 \times 2$ matrix of the type
$A =
\normalbaselineskip=15pt
\pmatrix{
a & b \cr
c & d \cr
}$,
$a \in M_1 \otimes M_N ({\gi C})$,
$b \in M ({\gi C}^2, {\gi C}) \otimes M_N ({\gi C})$,
$c \in M ({\gi C}, {\gi C}^2) \otimes M_N ({\gi C})$,
$d \in M_2 ({\gi C}) \otimes M_N ({\gi C})$. Note that
$A$ belongs to the commutant of
$(\ul H_l, \ul D_l, \ul \chi_l)$ iff one has
$b = c = 0$ and
$a = 1 \otimes \Gamma_N$,
$d = \bun \otimes \Gamma_N$ with
$\Gamma_N$ an arbitrary function of
$M_e$.
}

\item The commutant of the
$K$-cycle
$(\ul H_q, \ul D_q, \ul \chi_q)$ of
${\bf B}_{\rm chrom}$ coincides with
$\int (D_q, D_q) \otimes \bun_3$.

\item The commutant of the $K$-cycle
$(\ul H_l, \ul D_l, \ul \chi_l)$ of
${\bf B}_{\rm chrom}$ coincides with
$\int (D_l, D_l)$.

\end{enumerate}
}

\noindent \ul{Proof:} \begin{enumerate}

\item If
$\ul S$ commutes with
$\ul D_q$ and
$\ul \pi_q ({\bf A}_{\rm ew})$, it is of the form
$S \otimes T$,
$T \in M_N ({\gi C})$,
$S$ commuting with
$D_q$ and
$\pi_q ({\bf A}_{\rm ew})$.
$S =
\normalbaselineskip=18pt
\pmatrix{
a & b \cr
c & d \cr
}$,
$a, b, c, d \in M_2 ({\gi C}) \otimes M_N ({\gi C})$ is then both
of the form~\rf{35} and fulfills for all
$p \in {\gi H}_{\rm diag}$,
$q \in {\gi H}$:
  \alpheqn
\bea{44a}
& & [a, p \otimes \bun_N] = 0,                              \\[2mm]
& & [d, q \otimes \bun_N] = 0, \label{44b}                  \\[2mm]
& & b (q \otimes \bun_N) - (p \otimes \bun_N) b = 0, \label{44c}
                                                            \\[2mm]
& & c (p \otimes \bun_N) - (q \otimes \bun_N) c = 0, \label{44d}
\eea
  \reseteqn
Now~\rf{44c}, resp.~\rf{44d}, with
$p = 0$ yield
$b (q \otimes \bun_N) = 0$, resp.
$(q \otimes \bun_N) c = 0$ whence
$b = c = 0$ since
$q \otimes \bun_N$ is invertible. Since
${\gi H}$ generates
$M_2 ({\gi C})$ linearly, \rf{44b} further implies that
$d = \bun \otimes \ul d$,
with
$\ul d \in M_N ({\gi C})$. Together these imply that
$l = m = 0$,
$f (\mu_u) = C_g (\wt \mu_d) C^{\star}$ being a multiple of
the identity: this then also holds for
$S$.

\item If
$S$ commutes with
$D_l$ and
$\pi_l ({\bf A}_{\rm ew})$, it is of the form~\rf{34}
and is contained in
\be{45}
\pi_l ({\bf A}_{\rm ew})' =
\left\{
\normalbaselineskip=18pt
\pmatrix{
\bun_1 \otimes X & 0                \cr
0                & \bun_2 \otimes Y \cr
}
; X, Y \in M_N ({\gi C})
\right\},
\ee
(cf.~\rs{s1}(i)). Both facts together imply
$k (\mu_e) = 0$, and
$h (\mu_e) = \delta = X = Y$, whence the claim.

\end{enumerate}


\section{Proposition.}\label{s4}

{\it We have that:

\begin{enumerate}

\item The positive elements of the commutant of the
$K$-cycle
$(\ul H_q \oplus \ul H_l, \ul D_q \oplus \ul D_l,
\ul \chi_q \oplus \ul \chi_l)$
of
${\bf A}_{\rm ew}$ are the operators of the form
\be{46}
\ul \Gamma =
\normalbaselineskip=18pt
\pmatrix{
\ul \Gamma_q & 0            \cr
0            & \ul \Gamma_l \cr
}
\ee
with
$\ul \Gamma_q \in \bun_2 \otimes \bun_N \otimes M_3 ({\gi C})^+$ and
$\ul \Gamma_l$ of the type
$\bun_2 \otimes \Gamma_N$, with
$\Gamma_N$ a positive function of
$M_e$.\footnote{
The
$2 \times 2$ matrix corresponds to the decomposition
$\ul H_q \oplus \ul H_l$. Note that, as a consequence, we should in
Ansatz~\rf{3a}, \rf{3b} plug in
$\ul \Gamma_q = \bun_2 \otimes \bun_N \otimes \Gamma_3$ with
$\Gamma_3 \in M^+ ({\gi C}^3)$, and
$\Gamma_l = \bun_2 \otimes \Gamma_N$ with
$\Gamma_N$ a positive function of
$M_e$.
}
\item Consequently
$\ul \Gamma$ in~\rf{46} belongs to the center of
$(\pi_l \oplus \pi_q) ({\bf A}_{\rm ew})$ iff
$\ul \Gamma_l = \lambda \bun_{H_l}$ and
$\ul \Gamma_q = \lambda \bun_{H_q}$ for some real
$\lambda$.

\item The positive elements of the commutant of the $K$-cycle
$(\ul H_q \oplus \ul H_l, \ul D_q \oplus \ul D_l,
\ul \chi_q \oplus \ul \chi_l)$ of
${\bf B}_{\rm chrom}$ are the operators of the form
\be{47}
\ul \Gamma' =
\normalbaselineskip=18pt
\pmatrix{
\ul \Gamma'_q & 0             \cr
0             & \ul \Gamma'_l \cr
},
\ee
where
$\ul \Gamma'_q = \Gamma'_q \otimes \bun_3$ and
$\ul \Gamma'_l = \Gamma'_l$, with
$\Gamma'_q$ positive as in~\rf{35} and
$\Gamma'_l$ positive as in~\rf{34}.

\item Consequently
$\ul \Gamma'$ in~\rf{47} belongs to the
center of
$(\pi_l \oplus \pi_q) ({\bf B}_{\rm chrom})$ iff
one has
$\ul \Gamma'_l = \lambda'' \bun_{H_l}$ and
$\ul \Gamma'_q = \lambda' \bun_{H_q}$ for
some real
$\lambda',\ \lambda''$.

\end{enumerate}
}

\noindent \ul{Proof:} \begin{enumerate}

\item (resp. (iii)):~\rs{s1}(i) together with~\rs{s3}(i)(ii)
(resp.~\rs{s3}(iii)(iv) and~\rs{s2}(i)(ii)),
leads to the the diagonal elements of the matrix~\rf{46}
(resp.~\rf{46}); and together with~\rs{s2}(iii), it leads to the
vanishing of its off-diagonal elements.

\item Asking simultaneously for
$\pi_q (p, q) =
\normalbaselineskip=18pt
\pmatrix{
p & 0 \cr
0 & q \cr
} \otimes \bun_N \otimes \bun_3 \in
\bun_2 \otimes \bun_N \otimes M_3 ({\gi C})^+$ and
$\pi_l (p, q) =
\normalbaselineskip=18pt
\pmatrix{
p & 0 \cr
0 & q \cr
} \otimes \bun_N = \bun_2 \otimes \Gamma_N$ requires
$\Gamma_N = \lambda \bun_N$ and
$p = q = \lambda$ for some real
$\lambda$.

\item Asking simultaneously for
$\pi_l (p', m) = p' \bun_2 \otimes \bun_N = \Gamma'_l =
\normalbaselineskip=18pt
\pmatrix{
h (\mu_e) & 0      & k (\mu_e) \cr
0         & \delta & 0         \cr
k (\mu_e) & 0      & h (\mu_e) \cr
}$ and
$\pi_q (p', m) = \bun_2 \otimes \bun_N \otimes m = \ul \Gamma'_q =
\normalbaselineskip=18pt
\pmatrix{
f (\mu_u) & 0 & l (\mu_u) & 0 \cr
0 & g \left( \wt \mu_d \right) &
0 & m \left( \wt \mu_d \right) C^{\star} \cr
l (\mu_u) & 0 & f (\mu_u) & 0 \cr
0 & C m \left( \wt \mu_d \right) &
0 & C g \left( \wt \mu_d \right) C^{\star} \cr
} \otimes \bun_3$\break implies
vanishing of
$k (\mu_e)$,
$l (\mu_u)$ and
$m \left( \wt \mu_d \right)$, and equality of
$h (\mu_e)$,
$\delta$ to
$\lambda' \bun_N$ and of
$f (\mu_u)$,
$g \left( \wt \mu_d \right)$ to
$\lambda'' \bun_N$.

\end{enumerate}

Knowing the commutants of the
$K$-cycle
$(\ul H_q \oplus \ul H_l, \ul D_q \oplus \ul D_l,
\ul \chi_q \oplus \ul \chi_l)$ of
${\bf A}_{\rm ew}$ and of
${\bf B}_{\rm chrom}$ we know the general form of the new scalar
products~\rf{3a}, \rf{3b}. We can thus proceed to our computation of
the Yang-Mills action. We begin with the electroweak sector. We first
calculate the projection
$P_2$ key to the computation of quantum two-forms,  this will enable
us to compute the electroweak curvature corresponding to the new
scalar product~\rf{3a}, and the new form of the electroweak Yang-Mills
action (differing slightly  from that we had with the scalar
product~\rf{2}, the deviation being negligible in the valid
approximation where all fermion masses are neglected against that of
the top quark). We then compute the same objects for the
chromodynamics sector for which the projection
$P_2$ reduces to its spatial tensorial part due to the fact that the
gluons are of vectorial type, the Yang-Mills action needs however to
be computed anew. We then combine the electroweak and chromodynamics
sectors making use of the modular condition: this yields the total
Yang-Mills action, comprising terms identical to those of the
(bosonic) action of the traditional full standard model, however with
interesting constraints which we discuss at the end.

\bigskip

\noindent {\bf COMPUTATION OF THE ELECTROWEAK PROJECTION ${\gi P}_2$.}


\section{Proposition.}\label{s5} (canonical representant of quantum
two-forms).

{\it Consider the class
$\pi_{{\gi D}} (\Omega {\gb A}^2)$ modulo
$\pi_{{\gi D}} (\delta {\gb K}^1)$, indexed by
$(\lambda, \mu, {\bf q}, {\bf q}', {\bf Q}, {\bf Q}')$, consisting of
the direct sum of the elements
\be{48}
\left\{
\ba{ll}
\eta_{[2,0]} & =
\normalbaselineskip=18pt
\pmatrix{
\left[
\Bigl( \gamma (\lambda_k^i) \Bigr) -
(\ul X_k^i)
\right]
\otimes \bun_N & 0       \cr
0              & \left[
\Bigl( \gamma (\mu_k^i) \Bigr) -
(\ul Y_k^i)
\right]
\otimes \bun_N           \cr
}                                \\[6mm]
\eta_{[1,1]} & =
\normalbaselineskip=18pt
\pmatrix{
0              & {\gi M}^{\star}
\left[ i \ul \gamma ({\bf q}') \gamma^5 \otimes \bun_N \right]  \cr
\left[ i \ul \gamma ({\bf q}) \gamma^5 \otimes \bun_N \right]
{\gi M}        & 0                                              \cr
}                                \\[6mm]
\eta_{[0,2]} & =
\normalbaselineskip=18pt
\pmatrix{
{\gi M}^{\star} (\ul {\bf Q} \otimes \bun_N) {\gi M} & 0         \cr
0 & \ul {\bf Q}' \otimes \Sigma + i \ul {\bf Q}'' \otimes \Delta \cr
}
\ea
\right.
\ee
$$
\mbox{with}\
\left\{
\ba{l}
\left( \lambda_k^i \right) \in \Omega
 \left( {\bf M}, {\gi H}_{\rm diag} \right)^2 \\[2mm]
\left( X_k^i \right) \in C^{\infty}
 \left( {\bf M}, {\gi H}_{\rm diag} \right)   \\[2mm]
\left( \mu_k^i \right) \in \Omega
 \left( {\bf M}, {\gi H} \right)^2            \\[2mm]
\left( Y_k^i \right) \in C^{\infty}
 \left( {\bf M}, {\gi H} \right)              \\[2mm]
{\bf q}, {\bf q}' \in \Omega
 \left( {\bf M}, {\gi H} \right)^1            \\[2mm]
{\bf Q}, {\bf Q}', {\bf Q}'' \in C^{\infty}
 \left( {\bf M}, {\gi H} \right)
\ea
\right. , \
\left\{
\ba{ll}
\Sigma & = \dis \frac{1}{2}
 (\mu_u + \mu_d)     \\[2mm]
\Delta & = \dis \frac{1}{2}
 (\mu_u - \mu_d)
\ea
\right. , \
\left\{
\ba{ll}
\mu_u & = M_u M_u^{\star} \\[2mm]
\mu_d & = M_d M_d^{\star}
\ea
\right. \cdot
$$
and their leptonic reduction, where one fixes
$\left( \lambda_k^i \right)$,
$\left( \mu_k^i \right)$,
${\bf q}$,
${\bf q}'$,
${\bf Q}$,
${\bf Q}'$, and lets
$(X_k^i)$,
$(Y_k^i)$,
${\bf Q}''$ range through all possible values.

And take as scalar product the convex combination
$\alpha_q (\omega, \omega')_q + \alpha_l (\omega, \omega')_l$ of the
scalar products~\rf{3a} with the choices
$\ul \Gamma_q = \bun_2 \otimes \bun_N \otimes \Gamma_3$,
$\Gamma_3 \in M_3 ({\gi C})$ positive with
$Tr_3 \Gamma_3 = 3$, and
$\ul \Gamma_l = \bun_2 \otimes \Gamma_N$,
$\Gamma_N$ a positive function of
$M_e$ with
$Tr_N \Gamma_N = N$.\footnote{
As already noted these choices encompass the special cases
$\Gamma_3 = \bun_3$ and
$\Gamma_N = \bun_N$ corresponding to the previous Ansatz~\rf{2}.
}

Now the canonical representant of this class~\rf{38} (obtained by
projecting orthogonally parallel to
$\left( \pi_q \oplus \pi_l \right) (\delta {\gb K}^1)$ and indexed by
$(\lambda, \mu, {\bf q}, {\bf q}', {\bf Q}, {\bf Q}')$) is given as
follows: make in~\rf{38}:
\be{49}
\left\{
\ba{ll}
X     & = \left( \alpha_l + 6 \alpha_q \right)^{-1}
 N^{-1} L Q_{\rm diag}                           \\[2mm]
Y_k^i & = \left( \alpha_l + 3 \alpha_q \right)^{-1}
 (2 N)^{-1} L \ul {\bf Q}'_k\!{}^i               \\[2mm]
\ul {\bf Q}'' & = 0
\ea
\right. ,
\ee
where: }
\be{50}
L = Tr
\left[ 3 \alpha_q (\mu_u + \mu_d) + \alpha_l \Gamma_N \mu_e \right].
\ee

Note that these results are obtained from those of the previous
work~\cite{4b} by effecting the change
$\mu_e \to \Gamma_N \mu_e$. This will allow to exploit them easily.

\noindent \ul{Proof:} One finds the representative
$(\lambda, \mu, {\bf q}, {\bf q}', {\bf Q}, {\bf Q}')$ by asking the
direct sum of the element~\rf{48} and its leptonic reduction to be
orthogonal to the direct sum of all elements
\be{51}
\left\{
\ba{l}
\normalbaselineskip=18pt
\pmatrix{
(\ul S_k^i) \otimes \bun_N & 0   \cr
0 & (\ul T_k^i) \otimes \bun_N   \cr
} +
\normalbaselineskip=18pt
\pmatrix{
0 &  0                             \cr
0 &  i \ul {\bf R} \otimes \Delta  \cr
}                                     \\[6mm]
\mbox{where}\
\left\{
\ba{l}
\left( S_k^i \right)
\in C^{\infty} ({\bf M}, {\gi H}_{\rm diag})           \\[2mm]
\left( T_k^i \right) \in C^{\infty} ({\bf M}, {\gi H}) \\[2mm]
{\bf R} \in C^{\infty} ({\bf M}, {\gi H})
\ea
\right.
\ea
\right. ,
\ee
and their leptonic reductions: this amounts to:\footnote{
We here use the fact that the traces
$\tau_{{\gb D}_q}$, resp.
$\tau_{{\gb D}_l}$, are proportional to the spatial integrals of
$Tr_2 \otimes Tr_N \otimes Tr_3$ resp.
$Tr_2 \otimes Tr_N$ (note that the latter vanishes on
$\chi_q$-odd elements).
}
\be{52}
\ba{ll}
0 = & \alpha_q Re \left( Tr_2 \otimes Tr_N \otimes Tr_3 \right)
\biggl[
\left\{
\left[
(\ul X_k^i) \otimes \bun_N - {\gi M}^{\star}
\left( \ul {\bf Q} \otimes \bun_N \right) {\gi M}
\right]
\left[ (\ul S_k^i) \otimes \bun_N \right]
\right.                                                \\
& \left. \mkern -2mu +
\left[
(\ul Y_k^i) \otimes \bun_N -
\ul {\bf Q}' \otimes \Sigma - i \ul {\bf Q}'' \otimes \Delta
\right]
\left[
(\ul T_k^i) \otimes \bun_N +
i {\bf R} \otimes \Delta
\right]
\right\} \left( \bun_2 \otimes \bun_N \otimes \Gamma_3 \right)
\biggl]                                                \\[4mm]
& + \alpha_l Re \left( Tr_2 \otimes Tr_N \right)
\left[
\{ \mbox{leptonic reduction} \}
\left( \bun_2 \otimes \Gamma_N \right)
\right],
\ea
\ee
(we omitted the  purely imaginary term
$(Tr_2 \otimes Tr_N)
\Bigl[
i (\ul Y_k^i) {\bf R} \otimes \Delta -
i \ul {\bf Q}' {\bf R} \otimes \Sigma \Delta -
i \ul {\bf Q}'' (\ul T_k^i)$
$\otimes \Delta
\Bigr]$ vanishing under
$Re$). This yields (with independent vanishing of the three terms in
$(\ul S_k^i)$,
$(\ul T_k^i)$, and
${\bf R}$):
\be{53}
\ba{ll}
0 = & 3 \alpha_q \cdot Re
\left( Tr_2 \otimes Tr_N \right)
\biggl\{ \!
\left[
(\ul X_k^i) (\ul S_k^i)
\otimes \bun_N - {\gi M}^{\star}
\Bigl( (\ul S_k^i) \ul {\bf Q} \otimes \bun_N \Bigr) {\gi M}
\right]                                           \\[2mm]
    & +
\left[
(\ul Y_k^i) (\ul T_k^i)
\otimes \bun_N - \ul {\bf Q}'
(\ul T_k^i) \otimes \Sigma
\right] +
\left[
\ul {\bf Q}'' {\bf R} \otimes \Delta^2
\right] \! \!
\biggl\}                                          \\[3mm]
    & + \alpha_l Re \left( Tr_2 \otimes Tr_N \right)
\Bigl[
\{ \mbox{leptonic reduction} \}
\left( \bun_2 \otimes \Gamma_N \right)
\Bigr]
\ea
\ee

\begin{itemize}

\item[---] Vanishing of the
$(\ul S_k^i)$-term: we have, with
$(\ul X_k^i) =
\normalbaselineskip=18pt
\pmatrix{
\ol{\ul X} & 0     \cr
0          & \ul X \cr
}, (\ul S_k^i) =
\normalbaselineskip=18pt
\pmatrix{
\ol{\ul S} & 0     \cr
0          & \ul S \cr
}, \ul {\bf Q} =
\normalbaselineskip=18pt
\pmatrix{
\ol{\ul Q}_2   & \ul Q_1 \cr
- \ol{\ul Q}_1 & \ul Q_2 \cr
}:$
\eject

\be{54}
\ba{ll}
 0 & = Re (Tr_2 \otimes Tr_N)
\Biggl[
3 \alpha_q
\left\{
\normalbaselineskip=18pt
\pmatrix{
\ol{\ul X} \ol{\ul S} \bun_N & 0                  \cr
0                            & \ul X \ul S \bun_N \cr
} -
\normalbaselineskip=18pt
\pmatrix{
\ol{\ul S} \ol{\ul Q}_2 M_u^{\star} M_u &
 \ul S \ul Q_1 M_u^{\star} M_d              \cr
- \ol{\ul S} \ol{\ul Q}_1 M_d^{\star} M_u &
 \ul S \ul Q_2 M_d^{\star} M_d              \cr
}
\right\}                                       \\[6mm]
 &     \hphantom{\, + Re (Tr_2 \otimes Tr_N)}
+ \alpha_l
\left\{
\normalbaselineskip=18pt
\pmatrix{
0 & 0                  \cr
0 & \ul X \ul S \bun_N \cr
} -
\normalbaselineskip=18pt
\pmatrix{
0 & 0                             \cr
0 & \ul S \ul Q_2 M_e^{\star} M_e \cr
}
\right\}
\normalbaselineskip=18pt
\pmatrix{
\Gamma_N & 0        \cr
0        & \Gamma_N \cr
}
\Biggr]                                        \\[6mm]
 & = Re (Tr_2 \otimes Tr_N)
\Biggl[
3 \alpha_q
\left\{
\normalbaselineskip=18pt
\pmatrix{
\ol{\ul X} \ol{\ul S} \bun_N & 0                  \cr
0                            & \ul X \ul S \bun_N \cr
} -
\normalbaselineskip=18pt
\pmatrix{
\ol{\ul S} \ol{\ul Q}_2 M_u^{\star} M_u &
 \ul S \ul Q_1 M_u^{\star} M_d              \cr
- \ol{\ul S} \ol{\ul Q}_1 M_d^{\star} M_u &
 \ul S \ul Q_2 M_d^{\star} M_d              \cr
}
\right\}                                       \\[6mm]
 &     \hphantom{\, + Re (Tr_2 \otimes Tr_N)}
+ \alpha_l
\normalbaselineskip=18pt
\pmatrix{
0 & 0                    \cr
0 & \ul X \ul S \Gamma_N \cr
} -
\normalbaselineskip=18pt
\pmatrix{
0 & 0                                      \cr
0 & \ul S \ul Q_2 M_e^{\star} M_e \Gamma_N \cr
}
\Biggr]                                        \\[6mm]
 & = Re \biggl\{
3 \alpha_q
\left[
N (\ol{\ul X} \ol{\ul S} + \ul X \ul S) -
Tr \mu_u \cdot \ol{\ul S} \ol{\ul Q}_2 -
Tr \mu_d \cdot \ul S \ul Q_2
\right]                                        \\[3mm]
 &     \hphantom{\, + Re (Tr_2 \otimes Tr_N)}
+ \alpha_l
\left[
N \ul X \ul S - \ul S \ul Q_2 Tr (\Gamma_N \mu_e)
\right]
\biggr\}                                       \\[4mm]
 & =
3 \alpha_q
\left[
2 N \ul X \ul S -
(Tr \mu_u + Tr \mu_d) \ul S \ul Q_2
\right]
+ \alpha_l
\left[
N \ul X \ul S - \ul S \ul Q_2 Tr (\Gamma_N \mu_e)
\right]                                        \\[3mm]
 & = \ul S
\biggl[
3 \alpha_q
\left[
2 N \ul X -
(Tr \mu_u + Tr \mu_d) \ul Q_2
\right]
+ \alpha_l
\left[
N \ul X - \ul Q_2 Tr (\Gamma_N \mu_e)
\right]
\biggr]                                        \\[3mm]
 & = \ul S
\left[
(6 \alpha_q + \alpha_l) N \ul X - Tr
\Bigl( 3 \alpha_q (\mu_u + \mu_d) + \alpha_l \Gamma_N \mu_e \Bigr)
\ul Q_2
\right]                                        \\[3mm]
 & = \ul S
\left[
(6 \alpha_q + \alpha_l) N \ul X - L \ul Q_2
\right],
\ea
\ee
whence the relation first line in~\rf{49}.

\item[---] Vanishing of the
$(\ul T_k^i)$-term: we have:
\be{55}
\ba{lll}
 0 & = & Re (Tr_2 \otimes Tr_N)
\biggl\{
3 \alpha_q
\left[
(\ul Y_k^i) (\ul T_k^i)
\otimes \bun_N - \ul {\bf Q}' (\ul T_k^i)
\otimes \Sigma
\right]                              \\[2mm]
 & & + \alpha_l
\left[
(\ul Y_k^i) (\ul T_k^i)
\otimes \bun_N - \ul {\bf Q}' (\ul T_k^i)
\otimes \dis \frac{1}{2} \mu_e
\right]
\left( \bun_2 \otimes \Gamma_N \right)
\biggr\}                             \\[2mm]
 & = & Re
\biggl\{
3 \alpha_q
\left[
N \ul Y_k^i \ul T_i^k - \dis \frac{1}{2} Tr (\mu_u + \mu_d)
\ul {\bf Q}'_k\!{}^i \ul T_i^k
\right]                              \\[2mm]
 & & + \alpha_l
\left[
N \ul Y_k^i \ul T_i^k
- \dis \frac{1}{2} Tr (\Gamma_N \mu_e)
\ul {\bf Q}'_k\!{}^i \ul T_i^k
\right]
\biggr\}                             \\[2mm]
 & = & Re
\biggl\{
\ul T_i^k
\left[
(3 \alpha_q + \alpha_l) N \ul Y_k^i - \dis \frac{1}{2} Tr
\left[
3 \alpha_q (\mu_u + \mu_d) + \Gamma_N \mu_e
\right]
\ul {\bf Q}'_k\!{}^i
\right]
\biggr\}                             \\[3mm]
 & = & Re
\biggl\{
\ul T_i^k
\left[
(3 \alpha_q + \alpha_l) N \ul Y_k^i - \dis \frac{1}{2} L
\ul {\bf Q}'_k\!{}^i
\right]
\biggr\}
\ea
\ee
whence the relation second line in~\rf{49}.

\item[---] Vanishing of the
$\ul {\bf R}$-term:
$Re (Tr_2 \otimes Tr_N)
\left[
\ul {\bf Q}'' {\bf R} \otimes
\left( \alpha_q \Delta^2 + \alpha_l \mu_e^2 \right)
\right]$
vanishes for all
${\bf R}$ iff
$\ul {\bf Q}'' = 0$, the relation third line in~\rf{49}.

\end{itemize}

\pagebreak

\noindent {\bf CHROMODYNAMICS COMMUTANT OF THE INNER SPACE
${\gi K}$-CYCLE\break
AND CHRO\-MODYNAMICS PROJECTION
${\gi P}_2$.}


\section{Proposition.}\label{s6}

{\it Consider the class
$\pi_{{\gi D}} (\Omega {\gb B}^2)$ modulo
$\pi_{{\gi D}} (\delta {\gb K}^1)$, indexed by
$({\bf g}_k^i, {\bf f}')$, consisting of the direct sum of the
elements
\be{56}
\bun_{H_q} \otimes \bun_N \otimes
\left[
\ul \gamma ({\bf g}_k^i) + \ul X'_k\!{}^i
\right], \kern 1cm
\left\{
\ba{l}
({\bf g}_k^i) \in \Omega
\Bigl( {\bf M}, M_3 ({\gi C}) \Bigr)^2 \\[2mm]
\left( X'_k\!{}^i \right) \in C^{\infty}
\Bigl( {\bf M}, M_3 ({\gi C}) \Bigr)
\ea
\right. ,
\ee
and
\be{57}
\bun_{H_l} \otimes \bun_N \otimes
\left[
\ul \gamma ({\bf f}') + X'
\right], \kern 1cm
\left\{
\ba{l}
{\bf f}' \in \Omega ({\bf M}, {\gi C})^2 \\[2mm]
X' \in C^{\infty} ({\bf M}, {\gi C})
\ea
\right. ,
\ee
where one fixes
${\bf g}_k^i$,
${\bf f}'$, and lets
$X'_k\!{}^i$,
$X'$ range through all possible values. And take as scalar product the
sum of the scalar products~\rf{3b} with the choices
$\ul \Gamma'_q = \Gamma'_q \otimes \bun_{\rm chrom}$,
$\Gamma'_q$ positive in the commutant of
$D_q$, and
$\ul \Gamma'_l = \Gamma'_l$,
$\Gamma'_l$ positive in the commutant of
$D_l$.

The canonical representant of this class} (obtained by
projecting orthogonally parallel to
$(\pi_q \oplus \pi_l) (\delta {\gb K}^1)$ and indexed by
$\left( {\bf g}_k^i, {\bf f}' \right)$), {\it
is given as follows: make
$\left( X'_k\!{}^i \right) = 0$ in~\rf{56} and
$X' = 0$ in~\rf{57}.}
In other terms $P_2$ acts only on the space-time tensorial factor.

\noindent \ul{Proof:} Commutation of the generalized Dirac operator
with the action of the  chromodynamics algebra (cf.~\rf{26}) makes the
situation easy to handle. Since
${\bf B}_{\rm chrom}$ acts irreducibly on
${\gi C}_{\rm chrom}^3$, the quarkonic commutant is of the form
$\ul \Gamma'_q = \Gamma'_q \otimes \bun_{\rm chrom}$,
$\Gamma'_q$ in  the commutant of
$D_q$. And since
${\bf B}_{\rm chrom}$ acts on
${\gi C}_{\rm chrom}$ by scalars, the leptonic  commutant is of the
form
$\ul \Gamma'_l = \Gamma'_l$,
$\Gamma'_l$ in the commutant of
$D_l$. We recall that, by the commutation of
$(\ul \pi_q \oplus \ul \pi_l) ({\bf B}_{\rm chrom})$ with
$\ul D_q \oplus \ul D_l$,
$(\ul \pi_q \oplus \ul \pi_l)
\Bigl( (\Omega {\bf B}_{\rm chrom})^n \Bigr)$ vanishes for
$n \ge 1$, implying the simple situation
$\ul \pi_q \Bigl( (\Omega {\gb B})^2 \Bigr) =
\pi_D \Bigl( (\Omega {\gi A})^2 \Bigr)
\otimes \ul \pi_q ({\bf B}_{\rm chrom})$ and
$\ul \pi_l \Bigl( (\Omega {\gb B})^2 \Bigr) =
\pi_D \Bigl( (\Omega {\gi A})^2 \Bigr)
\otimes \ul \pi_l ({\bf B}_{\rm chrom})$, i.e. the form~\rf{56},
\rf{57} for the Hilbert space representant of formal two-forms.
Together with the product structure~\rf{4} of the traces,
this implies that, for the computation of
$P_2$ and of the chromodynamics part of the action,
$\Gamma_q$ and
$\Gamma_l$ enter only through their traces\footnote{
Thus we do not need to compute their precise form.
}
$Tr_N (\Gamma_q)$ and
$Tr_N (\Gamma_l)$. Hence
$P_2$ restricts to its space-time tensorial factor. Here is the
explicit calculation:
$P_2$ is found by asking the direct sum of~\rf{56} and \rf{57} with
fixed
${\bf g}_k^i$,
${\bf f}'$ to be fiberwise orthogonal to
$\pi_{{\gi D}} (\delta {\gb K}^1)$, i.e. to all direct sums of
\be{58}
\bun_2 \otimes \bun_N \otimes \ul G_k^i, \kern 1cm
G_k^i \in C^{\infty} \Bigl( {\bf M}, M_3 ({\gi C}) \Bigr),
\ee
and
\be{59}
\bun_2 \otimes \bun_N \otimes \ul F', \kern 1cm
F' \in C^{\infty} ({\bf M}, {\gi C}).
\ee
This amounts to vanishing under the Clifford trace, for all
$G_k^i$ and
$F'$, of the following
expressions:
\be{60}
Tr (\Gamma_q \cdot) \otimes Tr_3
\left\{
\bun_{H_q} \otimes \bun_N \otimes
\left(
\ul G_k^i
\left[
\ul \gamma ({\bf g}_k^i) + \ul X'_k\!{}^i
\right]
\right)
\right\}
= Tr (\Gamma_q) G_k^i X'_k\!{}^i,
\ee
\be{61}
Tr_7 \otimes Tr (\Gamma_q \cdot) \otimes Tr_3
\left\{
\bun_{H_l} \otimes \bun_N \otimes
\left(
F'
\left[
\ul \gamma ({\bf f}') + \ul X'
\right]
\right)
\right\}
= Tr (\Gamma_l) F' X',
\ee
this leading indeed to
$X'_k\!{}^i, = X' = 0$.

Our task is to calculate the Yang-Mills action for the compound
electroweak-chromodyna\-mics system, for which the curvature is
obtained via modular condition. We first ignore this subtlety and
compute the (unphysical) electroweak action. Once this is done; it
will be easy to adapt the computation to the full problem.

\bigskip

\noindent {\bf COMPUTATION OF THE ELECTROWEAK YANG-MILLS ACTION.}

\medskip

For the convenience of the reader we reproduce the expression of
the curvature~\cite{4b}: here is the quark component: with
$L = Tr
\left[ 3 \alpha_q (\mu_u + \mu_d) + \alpha_l \Gamma_N \mu_e \right]$,
${\bf f}$ the
$U (1)$-curvature,
${\bf h}_k^i$ the
$SU (2)$-curvature,
$\Phi$ the Higgs doublet, and
$V_{\Phi} = v_{\Phi}^2$ the Higgs potential:
\be{62}
\left\{
\ba{ll}
- \bthe_{q [2,0]} & =
\normalbaselineskip=18pt
\pmatrix{
\left[
\dis \frac{i}{2} \gamma ({\bf f})
\bun - (\alpha_l + 6 \alpha_q)^{-1}
N^{-1} L v_{\Phi} \bun
\right]
\otimes \bun_N & \kern -3cm  0  \cr
0              & \kern -3cm
\left[
\dis \frac{i}{2} \gamma ({\bf h}^\cdot_{\; \cdot})
- (\alpha_l + 3 \alpha_q)^{-1} (2 N)^{-1} L v_{\Phi} \bun
\right]
\otimes \bun_N \hfill           \cr
}                                   \\[9mm]
- \bthe_{q [1,1]} & =
\normalbaselineskip=18pt
\pmatrix{
0              & {\gi M}^{\star}
\left[
\gamma^5 \ul \gamma (i {\bf D} \Phi^{\star}) \otimes \bun_N
\right]                                                         \cr
\left[
\gamma^5 \ul \gamma (i {\bf D} \Phi) \otimes \bun_N
\right]
{\gi M}        & 0                                              \cr
}                                   \\[7mm]
- \bthe_{q [0,2]} & =
\normalbaselineskip=18pt
\pmatrix{
v_{\Phi} {\gi M}^{\star} {\gi M} & 0                            \cr
0                                & v_{\Phi} \bun \otimes \Sigma \cr
}
\ea
\right. ,
\ee
from which the leptonic component is obtained by suppression of the
first line and the first column of
$4 \times 4$ matrices, after the changes:
\be{63}
\left\{
\ba{cl}
{\gi M}    & =
\normalbaselineskip=18pt
\pmatrix{
\mu_u & 0      \cr
0     & \mu_d  \cr
} \to {\gi M} =
\normalbaselineskip=18pt
\pmatrix{
0 & 0      \cr
0 & \mu_e  \cr
}                        \\[6mm]
\Sigma & = \dis \frac{1}{2} (\mu_u + \mu_d) \to \Sigma =
\dis \frac{1}{2} \mu_e
\ea
\right. \cdot
\ee

\begin{itemize}

\item[---] Square of upper left corner: the leptonic part:
$$
\ba{lll}
  & \Biggl\{
\biggl[
\dis \frac{i}{2} \gamma ({\bf f}) & - (\alpha_l + 6 \alpha_q)^{-1}
N^{-1} L v_{\Phi}
\biggr]
\otimes \bun_N + v_{\Phi} {\gi M}^{\star} {\gi M}
\Biggr\}^2                                                  \\[2mm]
  & \hfill =
\biggl[ \!
& - \dis \frac{1}{4} \gamma ({\bf f})^2 + (\alpha_l + 6 \alpha_q)^{-2}
N^{-2} L^2 V_{\Phi}
\biggr]
\otimes \bun_N
+ V_{\Phi} {\gi M}^{\star} {\gi M} {\gi M}^{\star} {\gi M}  \\[4mm]
  & & - 2 (\alpha_l + 6 \alpha_q)^{-1}
N^{-1} L V_{\Phi} {\gi M}^{\star} {\gi M} + \mbox{terms linear in}\
\gamma ({\bf f})
\ea
$$
yielding after the substitutions~\rf{47} under
$\alpha_l\ tr_{\rm Clifford}
\otimes Tr_N (\Gamma_N \cdot)$:\footnote{
We use the fact that the normalized Clifford trace of
$\gamma ({\bf f})^2$ equals
$2 {\bf f}_{\mu \nu} {\bf f}^{\mu \nu}$.
}
$$
\ba{ll}
 \alpha_l
\biggl\{ \! \!
  & - \dis \frac{1}{2} N {\bf f}_{\mu \nu} {\bf f}^{\mu \nu} +
(\alpha_l + 6 \alpha_q)^{-2} N^{-1} L^2 V_{\Phi} +
Tr \left( \Gamma_N \mu_e^2 \right) V_{\Phi}          \\[2mm]
  & - 2 (\alpha_l + 6 \alpha_q)^{-1} N^{-1}
L Tr \left( \Gamma_N \mu_e \right) V_{\Phi}
\biggr\},
\ea
$$
and the quark part:
$$
\ba{lll}
   & \Biggl\{
\biggl[
\dis \frac{i}{2} \gamma ({\bf f}) \bun
& - (\alpha_l + 6 \alpha_q)^{-1} N^{-1} L v_{\Phi} \bun
\biggr]
\otimes \bun_N + v_{\Phi} {\gi M}^{\star} {\gi M}
\Biggr\}^2                                                  \\[2mm]
   & \hfill =
\biggl[ \!
& - \dis \frac{1}{4} \gamma ({\bf f})^2 \bun
+ (\alpha_l + 6 \alpha_q)^{-2} N^{-2} L^2 V_{\Phi} \bun
\biggr]
\otimes \bun_N
+ V_{\Phi} {\gi M}^{\star} {\gi M} {\gi M}^{\star} {\gi M}  \\[4mm]
   & & - 2 (\alpha_l + 6 \alpha_q)^{-1}
N^{-1} L V_{\Phi} {\gi M}^{\star} {\gi M} + \mbox{terms linear in}\
\gamma ({\bf f})
\ea
$$
yielding under
$\alpha_q\ tr_{\rm Clifford} \otimes Tr_2 \otimes Tr_N \otimes Tr_3
(\Gamma_3 \cdot)$:\footnote{
$Tr_2$ now denotes a trace in
$M_2 ({\gi C})$.
}
$$
\ba{ll}
 3 \alpha_q
\Bigl\{ \! \!
  & - N {\bf f}_{\mu \nu} {\bf f}^{\mu \nu} + 2
(\alpha_l + 6 \alpha_q)^{-2} N^{-1} L^2 V_{\Phi} +
Tr (\mu_u^2 + \mu_d^2) V_{\Phi}                 \\[2mm]
  & - 2 (\alpha_l + 6 \alpha_q)^{-1}
N^{-1} L Tr (\mu_u + \mu_d) V_{\Phi}
\Bigl\},
\ea
$$
combine to give:
$$
\ba{l}
  - \dis \frac{1}{2} N (\alpha_l + 6 \alpha_q)
{\bf f}_{\mu \nu} {\bf f}^{\mu \nu} +
(\alpha_l + 6 \alpha_q)^{-1} N^{-1} L^2 V_{\Phi} +
Tr
\left[
\alpha_l \Gamma_N \mu_e^2 +
3 \alpha_q (\mu_u^2 + \mu_d^2)
\right]
V_{\Phi}                                               \\[2mm]
  - 2 (\alpha_l + 6 \alpha_q)^{-1} N^{-1} L
Tr
\left[
\alpha_l \Gamma_N \mu_e +
3 \alpha_q (\mu_u + \mu_d)
\right]
V_{\Phi},
\ea
$$
i.e. (cf.~\rf{50}):
$$
- \frac{1}{2} N (\alpha_l + 6 \alpha_q)
{\bf f}_{\mu \nu} {\bf f}^{\mu \nu} -
(\alpha_l + 6 \alpha_q)^{-1} N^{-1} \ul L^2 V_{\Phi} +
Tr
\left[
\alpha_l \Gamma_N \mu_e^2 +
3 \alpha_q (\mu_u^2 + \mu_d^2)
\right]
V_{\Phi}
\leqno(*)
$$

\item[---] Contribution stemming from
$\bthe_{q [1,1]}$: we have:
$$
\ba{ll}
   (\bthe_{q [1,1]})^2 = &
\normalbaselineskip=18pt
\pmatrix{
{\gi M}^{\star}
\left[
\gamma^5 \ul \gamma (i {\bf D} \Phi^{\star}) \otimes \bun_N
\right]
\left[
\gamma^5 \ul \gamma (i {\bf D} \Phi) \otimes \bun_N
\right]
{\gi M} & \kern -3,5cm  0  \cr
0       & \kern -3,5cm
\left[
\gamma^5 \ul \gamma (i {\bf D} \Phi) \otimes \bun_N
\right]
{\gi M} {\gi M}^{\star}
\left[
\gamma^5 \ul \gamma (i {\bf D} \Phi^{\star}) \otimes \bun_N
\right]              \cr
}                            \\[6mm]
   & + {\gi M}^{\star}
\left[
\gamma^5 \ul \gamma (i {\bf D} \Phi^{\star}) \otimes \bun_N
\right]
\left[
\gamma^5 \ul \gamma (i {\bf D} \Phi) \otimes \bun_N
\right]
{\gi M},
\ea
$$
where the upper left corner can be replaced by its Hermitean part
$$
\frac{1}{2} {\gi M}^{\star}
\left\{
\left[
\ul \gamma ({\bf D} \Phi^{\star}), \ul \gamma ({\bf D} \Phi)
\right]_+ \otimes \bun_N
\right\}
{\gi M} = ({\bf D} \Phi_i) ({\bf D} \Phi^i) {\gi M}^{\star} {\gi M},
$$
also equal to the hermitean part of an order-permuted lower right
corner.

\noindent We thus get under
$\alpha_l\ tr_{\rm Clifford} \otimes Tr_2 \otimes Tr_N
(\Gamma_N \cdot)$ the leptonic part:
$$
2 \alpha_l Tr (\Gamma_N \mu_e) \cdot
({\bf D} \Phi_i) ({\bf D} \Phi^i),
$$
and under
$\alpha_q\ tr_{\rm Clifford} \otimes Tr_2 \otimes Tr_N \otimes Tr_3
(\Gamma_3 \cdot)$. The quark part:
$$
6 \alpha_q Tr (\mu_u + \mu_u) \cdot
({\bf D} \Phi_i) ({\bf D} \Phi^i),
$$
combining to give (cf.~\rf{50}):
$$
2 L ({\bf D} \Phi_i) ({\bf D} \Phi^i),
\leqno(**)
$$

\item[---] Square of lower right corner: the leptonic part, with the
substitutions~\rf{47}
$$
\ba{lll}
   & \Biggl\{
\biggl[
\dis \frac{i}{2} \gamma
({\bf h}^\cdot_{\; \cdot}) & -
(\alpha_l + 3 \alpha_q)^{-1}
(2 N)^{-1} L v_{\Phi} \bun
\biggr]
\otimes \bun_N + v_{\Phi} \bun \otimes \Sigma
\Biggr\}^2                                                 \\[2mm]
   & \hfill =
& - \dis \frac{1}{4}
\Bigl(
\gamma ({\bf h}_k^i) \gamma ({\bf h}_j^k)
\Bigr)
\otimes \bun_N + (\alpha_l + 3 \alpha_q)^{-2}
(2 N)^{-2} L^2 V_{\Phi} \bun
\otimes \bun_N + V_{\Phi} \bun
\otimes \dis \frac{1}{4} \mu_e^2                           \\[2mm]
   & & - 2 (\alpha_l + 3 \alpha_q)^{-1}
(2 N)^{-1} L V_{\Phi} \bun \otimes \dis \frac{1}{2} \mu_e
+ \mbox{terms linear in}\
\gamma ({\bf f})
\ea
$$
yielding under $\alpha_l\ tr_{\rm Clifford} \otimes Tr_2
\otimes Tr_N (\Gamma_N \cdot)$:\footnote{
We use the fact that the normalized Clifford trace of
$\bigl(
\gamma ({\bf h}_k^i) \gamma ({\bf h}_j^k)
\bigr)$ equals
${\bf h}_{\mu \nu}^1 {\bf h}_1^{\mu \nu}$.
}
$$
\ba{l}
   \alpha_l
\biggl\{
- \dis \frac{1}{4} N {\bf h}_{\mu \nu}^s {\bf h}_s^{\mu \nu} +
(\alpha_l + 3 \alpha_q)^{-2} (2 N)^{-1} L^2 V_{\Phi} +
\dis \frac{1}{2} Tr (\Gamma_N \mu_e^2) V_{\Phi}       \\[2mm]
   \hphantom{ \alpha_l \Bigl\{ } \!
- 2 (\alpha_l + 3 \alpha_q)^{-1}
(2 N)^{-1} L Tr (\Gamma_N \mu_e) \cdot V_{\Phi}
\biggr\}
\ea
$$
and the quark part:
$$
\ba{lll}
   & \Biggl\{
\biggl[
\dis \frac{i}{2} \gamma
({\bf h}^\cdot_{\; \cdot}) & -
(\alpha_l + 3 \alpha_q)^{-1}
(2 N)^{-1} L v_{\Phi} \bun
\biggr]
\otimes \bun_N + v_{\Phi} \bun \otimes \Sigma
\Biggr\}^2                                                 \\[2mm]
   & \hfill =
& - \dis \frac{1}{4}
\Bigl(
\gamma ({\bf h}_k^i) \gamma ({\bf h}_j^k)
\Bigr)
\otimes \bun_N + (\alpha_l + 3 \alpha_q)^{-2}
(2 N)^{-2} L^2 V_{\Phi} \bun
\otimes \bun_N + V_{\Phi} \bun \otimes \dis \frac{1}{4}
(\mu_u + \mu_d)^2                                          \\[2mm]
   & & - 2 (\alpha_l + 3 \alpha_q)^{-1}
(2 N)^{-1} L V_{\Phi} \bun \otimes \dis \frac{1}{2} (\mu_u + \mu_d)
+ \mbox{terms linear in}\
\gamma ({\bf f}),
\ea
$$
yielding under
$\alpha_q\ tr_{\rm Clifford} \otimes Tr_2 \otimes Tr_N \otimes Tr_3
(\Gamma_3 \cdot):^{10}$
$$
\ba{l}
   3 \alpha_q
\biggl\{
- \dis \frac{1}{4} N {\bf h}_{\mu \nu}^s {\bf h}_s^{\mu \nu} +
(\alpha_l + 3 \alpha_q)^{-2} (2 N)^{-1} L^2 V_{\Phi} +
\dis \frac{1}{2} Tr \left[ (\mu_u + \mu_d)^2 \right] V_{\Phi} \\[2mm]
   \hphantom{ 3 \alpha_q \Bigl\{ } \!
- 2 (\alpha_l + 3 \alpha_q)^{-1}
(2 N)^{-1} L Tr (\mu_u + \mu_d) \cdot V_{\Phi}
\biggr\}
\ea
$$
combine to give:
$$
\ba{l}
  - \dis \frac{1}{4} (\alpha_l + 3 \alpha_q)
N {\bf h}_{\mu \nu}^s {\bf h}_s^{\mu \nu} -
(\alpha_l + 3 \alpha_q)^{-1} (2 N)^{-1} L^2 V_{\Phi} +
\dis \frac{1}{2} Tr
\left[
\alpha_l \Gamma_N \mu_e^2 + 3 \alpha_q (\mu_u + \mu_d)^2
\right]
V_{\Phi}                                                  \\[2mm]
  - 2 (\alpha_l + 3 \alpha_q)^{-1}
(2 N)^{-1} L Tr
\left[
\alpha_l \Gamma_N \mu_e + 3 \alpha_q (\mu_u + \mu_d)
\right]
V_{\Phi}
\ea
$$
i.e. (cf.~\rf{50}):
$$
\ba{l}
   - \dis \frac{1}{4} (\alpha_l + 3 \alpha_q)
N {\bf h}_{\mu \nu}^s {\bf h}_s^{\mu \nu}          \\[2mm]
   - (\alpha_l + 3 \alpha_q)^{-1} (2 N)^{-1} L^2 V_{\Phi} +
\dis \frac{1}{2} Tr
\left[
\alpha_l \Gamma_N \mu_e^2 + 3 \alpha_q (\mu_u + \mu_d)^2
\right]
V_{\Phi}
\ea
\leqno(***)
$$

Collecting the terms ($\ast$), ($\ast \ast$), and ($\ast \ast \ast$)
yields the electroweak contribution.

\end{itemize}


\section{Proposition.}\label{s7}

{\it The electroweak Yang-Mills action equals\footnote{
We recall that
$\ul \Gamma_l = \bun_2 \otimes \Gamma_N$,
$\Gamma_N M_e Tr_N \Gamma_N = N$.
}
\be{64}
\ba{ll}
 Y M_{\rm ew} =
  & - \dis \frac{1}{2} (\alpha_l + 6 \alpha_q)
N {\bf f}_{\mu \nu} {\bf f}^{\mu \nu}
- \dis \frac{1}{4} (\alpha_l + 3 \alpha_q)
N {\bf h}_{\mu \nu}^s {\bf h}_s^{\mu \nu}    \\[3mm]
  & + 2 L ({\bf D} \Phi_j) ({\bf D} \Phi^j) +
K (\Phi_i \Phi^i - 1)^2,
\ea
\ee
Where
$\alpha_l$ and
$\alpha_q$ are positive constants, and where
\be{65}
\left\{
\ba{lll}
   L & = & Tr
\left[
\alpha_l \Gamma_N \mu_e + 3 \alpha_q (\mu_u + \mu_d)
\right]                                               \\[1mm]
   K & = & \dis \frac{3}{2} Tr
\left[
\alpha_l \Gamma_N \mu_e^2 + 3 \alpha_q (\mu_u^2 + \mu_d^2)
\right]                                               \\[3mm]
   & & + 3 \alpha_q Tr (\mu_u + \mu_d) -
\left[
2^{-1} (\alpha_l + 3 \alpha_q)^{-1} +
(\alpha_l + 6 \alpha_q)^{-1}
\right]
N^{-1} L^2
\ea
\right. .
\ee
}

\bigskip

\noindent {\bf COMPUTATION OF THE CHROMODYNAMICS YANG-MILLS ACTION.}

\medskip

We recall the expression of the leptonic, resp. quark components of
the gluonic curvature one has, with
${\bf f}'$ and
${\bf g}_0$
$U (1)$-curvature-two-forms, and
${\bf g}^a, a = 1, \dots, 8, a$
$S U (3)$-curvature-two-form (the
$\lambda_a$ are the eight Gell-Man matrices):$^{4,5}$
\be{66}
\theta_l = \frac{i}{2} \gamma ({\bf f}') \bun_2 \otimes \bun_N,
\ee
resp.
\be{67}
\theta_q = \frac{i}{2} \gamma ({\bf g}_0) \bun_2 \otimes \bun_N
\otimes \bun_3 +
\frac{i}{2} \gamma ({\bf g}^a) \bun_2 \otimes \bun_N
\otimes \frac{\lambda_a}{2} \cdotp
\ee
The gluonic new scalar product~\rf{3b} is given as
$Tr (\ul \Gamma \cdot)$,
$\ul \Gamma' =
\normalbaselineskip=18pt
\pmatrix{
\ul \Gamma'_q & 0             \cr
0             & \ul \Gamma'_l \cr
}$, (cf.~\rf{46}), with (cf.~\rf{34}, \rf{35}) here reproduced (in
shorthand) for the convenience of the reader:
$$
\ul \Gamma'_l =
\normalbaselineskip=18pt
\pmatrix{
h (\mu_e) & 0      & k (\mu_e) \cr
0         & \delta & 0         \cr
k (\mu_e) & 0      & h (\mu_e) \cr
},
\leqno(34)
$$
$$
\ul \Gamma'_q =
\normalbaselineskip=18pt
\pmatrix{
f (\mu_u) & 0 & l (\mu_u) & 0 \cr
0 & g \left( \wt \mu_d \right) &
0 & m \left( \wt \mu_d \right) C^{\star} \cr
l (\mu_u) & 0 & f (\mu_u) & 0 \cr
0 & C m \left( \wt \mu_d \right) &
0 & C g \left( \wt \mu_d \right) C^{\star} \cr
}
\otimes \bun_3.
\leqno(35)
$$
Recalling the formula:
\be{68}
Tr_3
\left\{
\left[
M \otimes \bun_3 + N^a \otimes \frac{\lambda_a}{2}
\right]^2
\right\}
= 3 M^2 + \frac{1}{2} N^a N_a,
\ee
$M$ and
$N^a$,
$a = 1, \dots, 8$,
$p \times p$ matrices, and since
\be{69}
\left\{
\ba{ll}
  \theta_l^2 & = - \dis \frac{1}{4} \gamma ({\bf f}')^2
\bun_2 \otimes \bun_N,                                   \\[2mm]
  \theta_q^2 & = -
\left[
\dis \frac{3}{4} \gamma ({\bf g}_0)^2 +
\dis \frac{1}{8} \gamma ({\bf g}^a) \gamma ({\bf g}_a)
\right]
\bun_2 \otimes \bun_N \otimes \bun_3
\ea
\right. ,
\ee
we have:
\be{70}
\ba{l}
  (Tr_2 \otimes Tr_N)
\left[ \ul \Gamma'_l \theta_l^2 \right] +
(Tr_2 \otimes Tr_N \otimes Tr_3)
\left[ \ul \Gamma'_q \theta_a^2 \right]               \\[2mm]
  \kern 0,9cm  = - \dis \frac{1}{4} \gamma ({\bf f}')^2 \cdot
(Tr_2 \otimes Tr_N) \ul \Gamma'_l - \dis \frac{1}{4} \gamma
\left[
3 \gamma ({\bf g}_0)^2 +
\dis \frac{1}{2} \gamma ({\bf g}^a) \gamma ({\bf g}_a)
\right]
\cdot (Tr_2 \otimes Tr_N \otimes Tr_3) \ul \Gamma'_q  \\[2mm]
  \kern 0,9cm  = - \dis \frac{1}{4} \gamma ({\bf f}')^2 \cdot
Tr_N [2 h (\mu_e) + \delta] - \dis \frac{1}{4}
\left[
3 \gamma ({\bf g}_0)^2 +
\dis \frac{1}{2} \gamma ({\bf g}^a) \gamma ({\bf g}_a)
\right]
\cdot Tr_N [2 f (\mu_u) + 2 g (\wt \mu_d)],
\ea
\ee
whose Clifford trace is then the sought gluonic Yang-Mills
action:


\section{Proposition.}\label{s8}

{\it The chromodynamics Yang-Mills action equals}

\be{71}
\ba{ll}
  Y M_{\rm chrom} = & - Tr_N [ h (\mu_e) + \delta / 2]
{\bf f}'_{\mu \nu} {\bf f}'{}^{\mu \nu}             \\[2mm]
                    & - Tr_N
\left[ f (\mu_u) + g (\wt \mu_d) \right]
\left[ 3 {\bf g}_{0 \mu \nu} {\bf g}_0^{\mu \nu}
+ \dis \frac{1}{2} {\bf g}_{\mu \nu}^a {\bf g}_a^{\mu \nu}
\right]
\ea
\ee
where
$h$,
$f$, and
$g$ are positive real functions.

\bigskip

\noindent {\bf MODULAR COMBINATION OF THE ELECTROWEAK AND
CHROMODYNAMICS SECTORS. THE FULL YANG-MILLS ACTION.}

\medskip

We now combine our results~\rs{s7} and~\rs{s8} using the modular
condition coalescing the unwanted three
$U (1)$-gauge groups into a single one. At the level of connexion the
modular condition amounts to the identifications:
\be{72}
\left\{
\ba{cl}
{\bf a}'  & = {\bf a}                  \\[1mm]
{\bf c}^0 & = - \dis \frac{1}{3} {\bf a}
\ea
\right. ,
\ee
\be{73}
\left\{
\ba{cl}
{\bf f}'  & = {\bf f}                    \\[1mm]
{\bf g}^0 & = - \dis \frac{1}{3} {\bf f}
\ea
\right. .
\ee
Adding the electroweak action~\rf{64} to the the gluonic
action~\rf{71} with these identifications then gives the total
Yang-Mills action:
\be{74}
\ba{ll}
 Y M =
  & - \dis \frac{1}{2}
\Biggl\{
N (\alpha_l + 6 \alpha_q) + Tr_N
\left[
2 h (\mu_e) + \delta + \dis \frac{2}{3}
\left[ f (\mu_u) + g (\wt \mu_d) \right]
\right] \!
\Biggr\}
{\bf f}_{\mu \nu} {\bf f}^{\mu \nu}           \\[3mm]
  & - \dis \frac{1}{4} N
(\alpha_l + 3 \alpha_q) {\bf h}_{\mu \nu}^s {\bf h}_s^{\mu \nu}
- \dis \frac{1}{2} Tr_N
\Bigl[ f (\mu_u) + g (\wt \mu_d) \Bigr]
{\bf g}_{\mu \nu}^a {\bf g}_a^{\mu \nu}       \\[5mm]
  & + 2 L ({\bf D} \Phi_j) ({\bf D} \Phi^j)
+ K (\Phi_i \Phi^i - 1)^2.
\ea
\ee


\section{Proposition.}\label{s9}

{\it The The bosonic action of the Connes-Lott version of the standard
model is of the form
\be{75}
Y M =
- A {\bf g}_{\mu \nu}^a {\bf g}_a^{\mu \nu}
- B {\bf f}_{\mu \nu} {\bf f}^{\mu \nu}
- \frac{1}{4} C {\bf h}_{\mu \nu}^s {\bf h}_s^{\mu \nu}
+ 2 L {\bf D}_{\mu} \Phi {\bf D}^{\mu} \Phi^j
+ K (\Phi_i \Phi^i - 1)^2 ,
\ee
where
\be{76}
\left\{
\ba{ll}
  A & = \dis \frac{1}{2} Tr_N
\left[ f (\mu_u) + g (\wt \mu_d) \right]                     \\[2mm]
  B & = \dis \frac{1}{2}
\Biggl\{
N (\alpha_l + 6 \alpha_q) + Tr_N
\left[
2  h (\mu_e) + \delta + \dis \frac{2}{3}
\left[ f (\mu_u) + g (\wt \mu_d) \right]
\right]
\!
\Biggr\}                                                     \\[4mm]
  C & = N (3 \alpha_q + \alpha_l)                            \\[3mm]
  L & = Tr
\Bigl[
\alpha_l \Gamma_N \mu_e + 3 \alpha_q (\mu_u + \mu_d)
\Bigr]                                                       \\[2mm]
  K & = \dis \frac{3}{2} Tr
\Bigl[
\alpha_l \Gamma_N \mu_e^2 + 3 \alpha_q (\mu_u^2 + \mu_d^2)
\Bigr]                                                       \\[3mm]
    & \hphantom{
K = \dis \frac{3}{2} Tr \alpha_l \Gamma_N \mu_e^2
            } \kern -3,3mm
+ 3 \alpha_q Tr (\mu_u + \mu_d) -
\left[
2^{-1} (\alpha_l + 3 \alpha_q)^{-1} + (\alpha_l + 6 \alpha_q)^{-1}
\right]
N^{-1} L^2
\ea
\right. ,
\ee
with
$\alpha_l, \alpha_q \ge 0$,
$\Gamma_N$ a positive function of
$\mu_e = M_e^2$ such that
$Tr_N \Gamma_N = N$, and
$f$ and
$g$ positive functions. The coupling constant is in the center iff
(a):
$\alpha_l = \alpha_q$~; (b):
$h (\mu_e) = \delta = \lambda' \bun_N$ for some
$\lambda' \ge 0$~; (c):
$f (\mu_u) = g (\wt \mu_d) = \lambda'' \bun_N$ for some
$\lambda'' \ge 0$.
}

\medskip

\noindent (the last claim follows from~\rs{s3}(ii) (requiring
$\alpha_q \bun_2 \otimes \bun_N \otimes \bun_3 =
\alpha_l \bun_2 \otimes \bun_N$) and~\rs{s3}(iv)).

Setting
$\alpha_l = \rho \frac{1 + x}{2}$,
$\alpha_q = \rho \frac{1 - x}{2}$,
$\rho > 0$,
$-1 \le x \le 1$,
$\frac{1}{2 \rho} Tr_N
\left[ f (\mu_u) + g (\wt \mu_d) \right] = y$,
$\frac{1}{2 \rho} Tr_N$ $[2 h (\mu_e) + \delta] = z$,
this reads:
 \setcounter{equation}{75}
  \alpheqn
\be{76a}
\left\{
\ba{ll}
  A / \rho & = y                                      \\[2mm]
  B / \rho & = N \dis \frac{7 - 5 x}{4} + z
+ \dis \frac{2}{3} y                                  \\[4mm]
  C / \rho & = N (2 - x)                              \\[2mm]
  L / \rho & = \dis \frac{1}{2} Tr
\Bigl[
(1 + x) \Gamma_N \mu_e + 3 (1 - x) (\mu_u + \mu_d)
\Bigr]                                                \\[3mm]
  K / \rho & = \dis \frac{3}{4} Tr
\biggl[
(1 + x) \Gamma_N \mu_e^2 + (1 - x)
\left[ 3 (\mu_u^2 + \mu_d^2) + 2 \mu_u \mu_d \right]
\biggr]                                                \\[2mm]
           & \hphantom{
K / \rho = \dis \frac{3}{4} Tr (1 + x) \Gamma_N \mu_e^2
            } \kern -7mm
- 3 \dis \frac{(5 - 3 x) (1 - x)}{2 (7 - 5 x) (2 - x)}
N^{-1} (L / \rho)^2
\ea
\right. .
\ee
  \reseteqn
with the coupling constants in the center iff
$x = 0$.

The expressions for
$L$ and
$K$ differ only from those obtained with the former
scalar pro\-duct~\rf{2} (besides the overall factor
$\rho$) by the factors
$\Gamma_N$ figuring in their first terms~\cite{4b} -- the latter
however without inference on the (very good\footnote{
The corresponding error on the tree-level computation of masses -- see
below -- is of the order of the  present error on the measurement of
the $W$ mass. The overall factor
$\rho$ drops out from the tree-level computation of masses.
}) approximation which consists in neglecting all fermion masses
against the top mass. This approximation leads to the values:
\be{77}
\left\{
\ba{cl}
  L / \rho & \cong \dis \frac{3}{2} (1 - x) m_t^2     \\[2mm]
  K / \rho & \cong \dis \frac{9}{4} (1 - x)
\left[
1 - \dis \frac{(5 - 3 x) (1 - x)}{2 (7 - 5 x) (2 - x)}
\right]
m_t^4
\ea
\right. .
\ee
The factors
$\Gamma_N$ are harmless: indeed we have
\be{78}
\Sup
\left\{
Tr (\Gamma_N^2) ; Tr \Gamma_N = N
\right\}
= \Sup
\left\{
\lambda_1^2 + \lambda_2^2 + \lambda_3^2 ;
\lambda_1 + \lambda_2 + \lambda_3 = N
\right\}
= 3,
\ee
thus, by the Schwartz inequality, one can neglect the terms:
\be{79}
\left\{
\ba{ll}
  | Tr (\Gamma_N \mu_e) |
& \le Tr (\Gamma_N^2)^{1/2} Tr (\mu_e^2) |^{1/2}
\le \sqrt 3 \, Tr (\mu_e^2) |^{1/2}                \\[2mm]
  | Tr (\Gamma_N \mu_e^2) |
& \le Tr (\Gamma_N^2)^{1/2} Tr (\mu_e^4) |^{1/2}
\le \sqrt 3 \, Tr (\mu_e^4) |^{1/2}
\ea
\right. ,
\ee

\bigskip

\noindent {\bf TREE-LEVEL COMPUTATIONS.}

\medskip

We now compare the Connes-Lott Lagrangian~\rf{75}
assorted with the covariant derivatives~\cite{4b}:
\be{80}
\left\{
\ba{ll}
  {\bf D}_{\mu}       & = \nabla_{\mu} + i
\left(
{\bf a}_{\mu} - {\bf b}_{\mu}^s \dis \frac{\tau_s}{2}
\right)                                                    \\[4mm]
  {\bf D}_{\mu}^{R l} & = \nabla_{\mu} - 2 i {\bf a}_{\mu} \\[3mm]
  {\bf D}_{\mu}^{L l} & = \nabla_{\mu}
- i a_{\mu} - i {\bf b}_{\mu}^s \dis \frac{\tau_s}{2}      \\[2mm]
  {\bf D}_{\mu}^{R u} & = \nabla_{\mu}
+ \dis \frac{4}{3} i {\bf a}_{\mu} - i {\bf c}_{\mu}^a
\dis \frac{\lambda_a}{2}                                   \\[2mm]
  {\bf D}_{\mu}^{R d} & = \nabla_{\mu}
- \dis \frac{2}{3} i {\bf a}_{\mu} - i {\bf c}_{\mu}^a
\dis \frac{\lambda_a}{2}                                   \\[2mm]
  {\bf D}_{\mu}^{L q} & = \nabla_{\mu}
+ \dis \frac{1}{3} i {\bf a}_{\mu}
- i {\bf b}_{\mu}^s \dis \frac{\tau_s}{2}
- i {\bf c}_{\mu}^a \dis \frac{\lambda_a}{2}
\ea
\right. ,
\ee
with the bosonic part of the traditional full standard
model:\footnote{
We recall the following relationships of constants in~\rf{81} with
masses and the weak angle: one has
$m_W = \frac{1}{2} v g_2$,
$m_H = \sqrt 2 \, \mu$, and
$\tang \theta_W = g_1 / g_2$.
}
\be{81}
\ba{ll}
 {\gb L}_{\rm gauge} + {\gb L}_{\rm Higgs} =
  & - \dis \frac{1}{4} G_{\mu \nu}^a G_a^{\mu \nu}
- \dis \frac{1}{4} B_{\mu \nu} B^{\mu \nu}
- \dis \frac{1}{4} W_{\mu \nu}^s W_s^{\mu \nu}        \\[2mm]
  & + (D_{\mu} \phi)^{\star} (D_{\mu} \phi)
+ \dis \frac{\mu^2}{v^2}
\left(
\phi^{\star} \phi - \dis \frac{v^2}{2}
\right)^2
\ea
\ee
assorted with the covariant derivatives:
\be{82}
\left\{
\ba{ll}
  D_{\mu}       & = \partial_{\mu}
- i \dis \frac{g_1}{2} B_{\mu}
- i g_2 W_{\mu}^s \dis \frac{\tau_s}{2}                     \\[4mm]
  D_{\mu}^{R l} & = \partial_{\mu} + i g_1 B_{\mu}          \\[3mm]
  D_{\mu}^{L l} & = \partial_{\mu}
+ i \dis \frac{g_1}{2} B_{\mu}
- i g_2 W_{\mu}^s \dis \frac{\tau_s}{2}                     \\[2mm]
  D_{\mu}^{R u} & = \partial_{\mu}
- 2 i \dis \frac{g_1}{3} B_{\mu}
- i g_3 G_{\mu}^a \dis \frac{\lambda_a}{2}                  \\[2mm]
  D_{\mu}^{R d} & = \partial_{\mu}
+ i \dis \frac{g_1}{3} B_{\mu}
- i g_3 G_{\mu}^a \dis \frac{\lambda_a}{2}                  \\[2mm]
  D_{\mu}^{L q} & = \partial_{\mu}
- i \dis \frac{g_1}{6} B_{\mu} - i g_2 W_{\mu}^s
\dis \frac{\tau_s}{2}
- i g_3 G_{\mu}^a \dis \frac{\lambda_a}{2}
\ea
\right.
\ee
Identification of the covariant derivatives~\rf{80} and \rf{82} is
synonymous with the identifications:
\be{83}
\left\{
\ba{l}
  {\bf c} = g_3 G                                       \\[2mm]
  \mbox{in components}\ {\bf c}_{\mu}^a = g_3 G_{\mu}^a,
\kern 2cm  a = 1, \dots , 8
\ea
\right. ,
\ee
\be{84}
\left\{
\ba{l}
  {\bf a} = - \dis \frac{1}{2} g_1 B     \\
  \mbox{in components}\ {\bf a}'_{\mu} =
- \dis \frac{1}{2} g_1 B_{\mu}
\ea
\right. ,
\ee
\be{85}
\left\{
\ba{l}
  {\bf b} = g_2 W                                      \\[2mm]
  \mbox{in components}\ {\bf b}_{\mu}^s = g_2 W_{\mu}^s,
\kern 2cm  s = 1, 2, 3.
\ea
\right.
\ee
implying:
\be{86}
\left\{
\ba{clr}
  {\bf f}_{\mu \nu}    & = - \dis \frac{1}{2} g_1 B_{\mu \nu}
= - \dis \frac{1}{2} g_2 \cos \theta_W B_{\mu \nu} \kern 2cm
&                                                           \\[2mm]
  {\bf h}_{\mu \nu}^s  & = g_2 W_{\mu \nu}^s ,
& s = 1, 2, 3,                                              \\[2mm]
  {\bf g}_{\mu \nu}^a  & = g {\gb G}_{\mu}^a ,
& a = 1, \dots, 8,
\ea
\right.
\ee
Assuming that
$\phi$ and
$\Phi$ differ by a constant (insensitive to multiplication of
$\phi$, resp.
$\Phi$, by  constants), the latter follows from comparison of the
fourth terms of~\rf{75} and \rf{81}, yielding :
\be{87}
\frac{v}{\sqrt 2} \Phi = \phi.
\ee
Inserting~\rf{83}, \rf{84} and~\rf{85} into~\rf{81} yields:
\be{88}
\ba{ll}
Y M =
  & - g_3^2 A \cdot G_{\mu \nu}^a G_a^{\mu \nu}
- g_1^2 \dis \frac{B}{4} B_{\mu \nu} B^{\mu \nu}
- \dis \frac{1}{4} g_2^2 C W_{\mu \nu}^s W_s^{\mu \nu}        \\[2mm]
  & + \dis \frac{4 L}{v^2} (D_{\mu} \phi)^{\star} (D^{\mu} \phi)
+ \dis \frac{4 K}{v^4}
\left(
\phi^{\star} \phi - \dis \frac{v^2}{2}
\right)^2 .
\ea
\ee
Comparison with~\rf{75} yields
\be{89}
4 g_3^2 A = g_1^2 B = g_2^2 C = \frac{4 L}{v^2}
= \frac{4 K}{\mu^2 v^2} \raise 2pt \hbox{,}
\ee
\be{90}
g_3 = \frac{1}{2} (C / A)^{1/2} g_2,
\ee
\be{91}
\frac{g_2^2}{g_1^2} = \tang^{- 2} \theta_W = \frac{B}{C}
\qquad  \mbox{whence}  \qquad
\sin^{2} \theta_W = \frac{C}{B + C}
\ee
\be{92}
v^2 g_2^2 = 4 m_W^2 = 4 \frac{L}{C}
\qquad  \mbox{whence}  \qquad
m_W = (L / C)^{1/2}
\ee
\be{93}
\mu^2 = K / L
\qquad  \mbox{whence}  \qquad
m_H = (2 K / L)^{1/2}.
\ee

Plugging into those relations
$A, B, C$ as in~\rf{76a} and
$K, L$ as in \rf{77} gives:


\section{Proposition.}\label{s10}

{\it We have the following tree-level evaluations:
\be{94}
g_3 = \frac{1}{2} (N (2 - x) / y)^{1/2} g_2
\ee
\be{95}
\sin^2 \theta_W = \frac{
N (2 - x)
}{
N \frac{3 (5 - 3 x)}{4} + z + \frac{2}{3} y
}
\ee
\be{96}
m_W \cong
\left[
\frac{3 (1 - x)}{2 N (2 - x)}
\right]^{1/2} m_t
\ee
and
\be{97}
m_H \cong \sqrt 3 \,
\left[
1 - \frac{ (5 - 3 x) (1 - x) }{ 2 (7 - 5 x) (2 - x) }
\right]^{1/2} m_t,
\ee
where
$\cong$ denotes approximation neglecting all fermion masses against
the Higgs mass. The real numbers
$x, y$, and
$z$ range independently in the respective intervals
$[-1, +1]$,
$[0, + \infty]$ and
$[0, - \infty]$.

The choice of the coupling constant in the center gives
}
\be{98}
\left\{
\ba{cl}
  m_t & = 2 m_W                                      \\[2mm]
  m_H & = \sqrt 3 \cdot \sqrt{(23 / 28)} \, m_t = 1.5698 m_t
\ea
\right. .
\ee

We now study the constraints affecting the above quantities. Since
$y$ is an arbitrary positive number,~\rf{94} contains no information.

\begin{itemize}

\item[---] Constraint on
$\sin^2 \theta_W$: since
$y, z \in [0, + \infty]$ one has
$\sin^2 \theta_W < \frac{4}{3} \frac{2 - x}{5 - 3 x} =
\frac{2}{3} \frac{m_t^2}{m_t^2 + m_W^2} \cdotp$

\item[---] Constraint on
$m_W / m_t$: continuous decreasing function of
$x$ varying from
$0$ to
$N^{- 1/2}$.

\item[---] Constraint on
$m_H / m_t$: continuous increasing function of
$x$ varying from
$\sqrt{(7 / 3)}$ to
$\sqrt 3$.

\end{itemize}

\noindent These are individual constraints. But we have also a
correlation due to the additional information obtained by eliminating
$x$ between~\rf{96} and \rf{97}: if we set:
\be{99}
S = \frac{2 N}{3} \left[ \frac{m_W}{m_t} \right]^{1/2} =
\frac{1 - x}{2 - x}
\qquad  \mbox{whence}  \qquad
x = \frac{2 S - 1}{S - 1}
\ee
we have
\be{100}
T = \frac{1}{3} \left[ \frac{M_H}{m_t} \right]^{1/2} =
1 - \frac{ (5 - 3 x) (1 - x) }{ 2 (7 - 5 x) (2 - x) } =
1 - \frac{ (5 - 3 x) }{ 2 (7 - 5 x) } S =
1 - \frac{1}{2} \frac{S + 2}{3 S + 2} S.
\ee


\section{Proposition.}\label{s11}

{\it The Connes-Lott version of the full standard model differs
in tree-approximation from the traditional standard model by the
following constraints relative to the weak angle and the ratios of
the top quark, resp. the Higgs boson mass to the
$W$-boson mass. Irrespective of the choice of coupling constant in the
$K$-cycle commutant we have the inequalities:
\be{101}
\sin^2 \theta_W \le \frac{2}{3} \frac{m_t^2}{m_t^2 + m_W^2}
\raise 2pt \hbox{\rm ,}
\ee
\be{102}
0 \le m_W / m_t \le N^{- 1/2},
\ee
\be{103}
\sqrt{(7 / 3)} \le m_W / m_t \le \sqrt 3.
\ee
Furthermore the ratios
$m_W / m_t$ and
$m_H / m_t$ determine each other as
follows: one has:
\be{104}
T = 1 - \frac{1}{2} \frac{S + 2}{3 S + 2} S
\qquad  \mbox{where}  \qquad
S = \frac{2 N}{3} \left[ \frac{m_W}{m_t} \right]^{1/2},
\qquad
T = \frac{1}{3} \left[ \frac{m_H}{m_t} \right]^{1/2},
\ee
where the values~\rf{98} corresponding to the coupling constant in the
$K$-cycle center are obtained for
$S = \frac{1}{2}$.
}

We conclude with the remark that the correlation~\rf{104} together
with the fact that the top has approximately twice the
$W$-mass leads to propose that the Higgs mass might lie around
$1.5698 m_t$.

\newpage

\section*{Appendix.}

Present notation versus notation in the companion paper~\cite{5a}
$$
\ba{ccll}
  \qquad \mbox{\cite{5a}} \qquad  & \mbox{this paper}\qquad &
                               &                 \\[2mm]
  x & \alpha_q \Gamma_3  & \qquad  (tr \Gamma_3 = 3)
               &    tr x = 3 \alpha_q     \\[2mm]
  y & \alpha_l \Gamma_N  &  \qquad \qquad\qquad (tr \Gamma_N = N)
               &    tr y = N \alpha_l     \\[2mm]
  r       & f (\mu_u)     & &                         \\[2mm]
  s       & g (\wt \mu_d) & &                         \\[2mm]
  u       & \delta        & &                         \\[2mm]
  v       & h (\mu_e)     & &                         \\[2mm]
  k       & l (\mu_u)     & &                         \\[2mm]
  p       & m (\wt \mu_d) & &                         \\[2mm]
  w       & k (\mu_e)     & &                         \\[1cm]
   C_{K M} & C             & &
\ea
$$

\end{document}